\documentclass[aps,pre,twocolumn,showpacs]{revtex4}
\usepackage{epsf}
\usepackage{amsmath,amssymb}

\newcommand{\Heff}{ \mathcal{H}_{\textrm{eff}} }
\newcommand{\HeffD}{ \Heff^{\textsc{(d)}} }
\newcommand{\HeffN}{ \Heff^{\textsc{(n)}} }
\newcommand{\Hint}{ H_{\textrm{int}} }
\newcommand{\HintD}{ \Hint^{\textsc{(d)}} }
\newcommand{\HintN}{ \Hint^{\textsc{(n)}} }
\newcommand{\Hext}{ H_{\textrm{ext}} }
\newcommand{\HextD}{ \Hext^{\textsc{(d)}} }
\newcommand{\HextN}{ \Hext^{\textsc{(n)}} }
\newcommand{\Gint}{ G_{\textrm{int}} }
\newcommand{\GintD}{ \Gint^{\textsc{(d)}} }
\newcommand{\GintN}{ \Gint^{\textsc{(n)}} }
\newcommand{\Ree}{ \Re\textrm{e}\, }
\newcommand{\Imm}{ \Im\textrm{m}\, }

\begin{document}

\title {Is the concept of the non-Hermitian effective Hamiltonian \\
relevant in the case of potential scattering?}

%
\received{11 June 2002} %
\published{26 February 2003 in: Phys. Rev. E \textbf{67}, 026215 (2003)}

\author{
Dmitry V. Savin$^{1,2}$,
Valentin V. Sokolov$^{2}$ and
Hans-J\"urgen Sommers$^1$ }
\affiliation{
$^1$Fachbereich Physik, Universit\"at-GH Essen, 45117 Essen,
Germany\\
$^2$Budker Institute of Nuclear Physics, 630090 Novosibirsk, Russia
}

\begin{abstract}
We examine the notion and properties of the non-Hermitian effective
Hamiltonian of an unstable system using as an example potential
resonance scattering with a fixed angular momentum. We present a
consistent self-adjoint formulation of the problem of scattering on a
finite-range potential, which is based on separation of the
configuration space into two segments, internal and external. The
scattering amplitude is expressed in terms of the resolvent of a
non-Hermitian operator $\mathcal{H}$. The explicit form of this
operator depends on both the radius of separation and the boundary
conditions at this place, which can be chosen in many different ways.
We discuss this freedom and show explicitly that the physical
scattering amplitude is, nevertheless, unique although not all
choices are equally adequate from the physical point of view.

The {\it energy-dependent} operator $\mathcal{H}$ should not be
confused with the non-Hermitian effective Hamiltonian $\Heff$ which
is usually exploited to describe interference of overlapping
resonances. We note that the simple Breit-Wigner approximation is as
a rule valid for any individual resonance in the case of few-channel
scattering on a flat billiard-like cavity, leaving no room for
non-trivial $\Heff$ to appear. The physics is appreciably richer in
the case of an open chain of $L$ connected similar cavities whose
spectrum has a band structure. For a fixed band of $L$ overlapping
resonances, the smooth energy dependence of $\mathcal{H}$ can be
ignored so that the constant $L\times L$ submatrix $\Heff$
approximately describes the time evolution of the chain in the energy
domain of the band and the complex eigenvalues of $\Heff$ define the
energies and widths of the resonances. We apply the developed
formalism to the problem of a chain of $L$ $\delta$ barriers whose
solution is also found independently in a closed form. We construct
$\Heff$ for the two commonly considered types of the boundary
conditions (Neumann and Dirichlet) for the internal motion. Although
the final results are in perfect coincidence, somewhat different
physical patterns arise of the trend of the system with growing
openness. Formation in the outer well of a short-lived doorway state
shifted in the energy is explicitly demonstrated together with the
appearance of $L\!-\!1$ long-lived states trapped in the inner part
of the chain.
\end{abstract}

\pacs{05.60.Gg, 03.65.Nk, 24.30.-v, 73.23.-b}
%

\maketitle

\section{Introduction}

To the best of our knowledge, the concept of the non-Hermitian
effective Hamiltonian appeared first in Feshbach's papers
\cite{Feshbach1958,Feshbach1962} in connection with the general
theory of resonance nuclear reactions \cite{Kapur1938,Wigner1947}
and, independently, in Liv{\v{s}}ic's study of open systems
\cite{Livsic1957,Livsic1973}. A typical atomic nucleus forms near an
excitation energy $E$ above the threshold of nucleon emission a rich
set of long-lived compound states with a very dense energy spectrum.
For relatively small excitation energies, these states manifest
themselves as narrow isolated resonances in collisions of nucleons
and target nuclei. The sharp energy dependence of the corresponding
cross sections near a given resonance is described by the universal
Breit-Wigner formula. Any smoother variations can be ignored in the
domain of an isolated resonance. At higher $E$, the resonance states
begin to overlap and strongly interfere. Nevertheless, one can still
neglect the smooth energy dependence within a group of close
interfering resonances. In this approximation, the propagation of the
unstable system created on the intermediate stage of the collision is
characterized by the resolvent of an energy-independent non-Hermitian
matrix \cite{Mahaux1969}. The latter describes in the time picture
the irreversible evolution of the excited intermediate state and can
therefore be interpreted as the effective Hamiltonian $\Heff$ whose
anti-Hermitian part, originating from on-shell self-energy
contributions, is responsible for decays into open channels. The
complex eigenvalues of the effective Hamiltonian determine the
energies and widths of the resonances. The method of the effective
Hamiltonian proved to be a success for describing many important
resonance phenomena, especially in the context of chaotic scattering
\cite{Verbaarschot1985,Sokolov1989,khimiki,
Rotter1991,Fyodorov1997,Dittes2000,Alhassid2000}.

It must be stressed that, similar to the Breit-Wigner formula, the
notion of the effective Hamiltonian is {\it local} in energy although
the general approaches \cite{Kapur1938,Wigner1947} work in much wider
energy intervals where smooth variations become important and should
already be taken into account. For the very concept of the effective
Hamiltonian to be consistent, the scales of resonant and smooth
variations must be appreciably different. Otherwise the resonance
behavior of scattering amplitudes will be distorted or even
completely destroyed. The smooth dependence influences the background
phases as well as the parameters of the resonances situated in
different energy domains. Such effects cannot be described by simply
enlarging the dimension of the matrix of the effective Hamiltonian.
Instead, a large energy-dependent matrix $\mathcal{H}(E)$ emerges
whose simple interpretation as a time-shift operator is no longer
valid. There is, generally, no one-to-one correspondence between the
complex energies of resonance states on the one hand and the
$E$-dependent eigenvalues of this large matrix on the other.

The current intensive studies, theoretical as well as experimental,
of chaotic scattering of a particle by an open two-dimensional cavity
(see, e.g., \cite{Stoeckmann1999} and references therein) renewed
interest in the Hamiltonian approach to the resonance scattering
theory and, in particular, to the concept and properties of the
effective non-Hermitian Hamiltonian
\cite{Akguc2001,Pichugin2001,Stoeckmann2002i}. Although the
scattering processes considered are purely potential, the theory is
formulated in close analogy with the formalism developed in
\cite{Feshbach1958,Feshbach1962,Kapur1938,Wigner1947} for nuclear
collisions. The configuration space is divided into two parts,
internal and external. The Feshbach projection technique is employed
to express the scattering matrix in the specific form that explicitly
reveals the sharp resonant energy dependence while all smoother
variations show up only indirectly via changes of the matrix
elements. Related is the problem of electromagnetic field
quantization in open optical cavities \cite{Hackenbroich2002}.

While the probability amplitudes of the physical processes are fixed
unambiguously, the projection procedure exploited is not unique.
There exists a rather wide freedom in choosing the surface of
separation as well as the boundary conditions (BCs) on this surface.
Therefore, the amplitudes of interest are expressed in terms of
quantities that depend on the details of the formalism. The
independence of the final results of the calculations is not, as a
rule, directly seen. Therefore, a certain caution is necessary to
avoid incorrect assertions. An unexpected dependence on the
(auxiliary) boundary condition at the cavity-lead interface was
found, for example, in \cite{Pichugin2001}. Moreover, poor agreement
with numerically calculated exact poles of the $S$ matrix was
revealed for poles distant from the real axis. These points, partly
attributed by the authors to numerical limitations, require further
clarification. Additional physical arguments should sometimes be
involved to reasonably restrict the freedom.

It is important to recognize that the density of levels in
billiard-like cavities is much lower than that of the many-body
nuclear systems. Actually, strong overlap and interference of
different resonance states are not, as a rule, possible in the single
two-dimensional cavities ordinarily considered \cite{Pichugin2001}.
The simple Breit-Wigner approximation is usually sufficient in this
case for any individual resonance (see, e.g.,
\cite{Alt1993,Alt1995}). Noticeable interference can occur only in
the rare event of accidental near degeneracy of resonances.
Experimental observation of such an interference of few resonances in
an open microwave cavity has recently been reported in
\cite{Persson2000}. Stronger overlap and a non trivial effective
Hamiltonian matrix can, however, appear when open chains of a number
of similar potential wells connected to each other are considered.
The energy spectrum has a band structure in this case with much
denser spectrum within a given band. A schematic model of such a kind
was considered in \cite{Sokolov1992}. A similar example was also
investigated in \cite{Dittes2000} in the framework of graph theory
\cite{Kottos1999}. Some general aspects of scattering in periodic
structures have also been considered in \cite{Barra1999}.

In this paper we examine the notion of the effective Hamiltonian in
potential scattering. The questions we are concerned with do not
depend on the regular or irregular character of the motion. Thus, we
restrict ourselves to the simplest case of the single-channel
$s$-wave scattering. The extension to higher partial waves is
straightforward. A consistent self-adjoint formulation of the problem
of scattering on a finite-size potential in terms of the internal and
external subproblems is presented in Sec.~II. The (one-dimensional)
$S$ matrix is expressed in terms of a non-Hermitian energy-dependent
operator ${\cal H}$ whose form, together with the form of the
associated $R$ function (Sec.~III), {\it depends} on the BC and the
radius $a$ of separation in the configuration space. Different BCs
define different representations of the internal and external parts
of the (unique) scattering wave function. The role of the fictitious
direct reflection at the separation point $a$ is also discussed.

In Sec.~IV the exactly solvable open Kronig-Penney model is
considered as an example of potential scattering with interfering
resonances. Closed expressions are found for both scattering $S(k)$
and staying-wave $K(k)$ functions as well as Wigner's function
$R(k)$. Their analytical properties in the complex $k$ and energy
planes are analyzed in detail. Hereafter we compare these exact
results with those obtained in the framework of the projection
formalism of the previous sections. We analyze two typical choices of
the BCs for the internal motion: the cases of Neumann and Dirichlet
BCs. In the second case, the internal problem corresponds to a closed
counterpart of the system under consideration that allows one to
follow the changes of the motion due to the interference of
resonances when the openness grows. The latter is in line with
numerous applications considered in the literature. Non-Hermitian
effective Hamiltonians for a fixed band of resonance states are built
up in both cases. We summarize our main findings in the concluding
Sec.~V.

\section{Separation of the Hilbert space}

The radial motion in the $s$-wave scattering is described by the
Schr\"odinger equation (we use the units $\hbar^2/2m=1$ throughout
the paper)
\begin{equation}
\left(-\frac{d^2}{dr^2} + U(r)\right)\chi(r) = k^2\chi(r)\
\label{Schr}
\end{equation}
with the boundary condition $\chi(0)=0$ at the origin. We use below
the specific method which, basically, goes back to Bloch's paper
\cite{Bloch1957} (see also \cite{Lane1966}). However, we present a
derivation which leads to interrelated boundary problems for internal
and external regions. Making use of Heaviside's step function one can
decompose the wave function into internal and external parts as
follows
\begin{equation}\label{psi1}
\chi(r) = u(r)\,\theta(a-r)+\phi(r)\,\theta(r-a)\,,
\end{equation}
where the functions $u$ and $\phi$ are supposed to be continuous at
the point $a$ together with their two first derivatives. The
partition radius $a$ can be chosen arbitrarily; we suggest only that
$U(r>a)\equiv 0$. The second derivative of the function (\ref{psi1})
reads
\begin{eqnarray}\label{psi''}
&\chi''(r)= u''(r)\,\theta(a-r) +
 \phi''(r)\,\theta(r-a) +          \nonumber \\
& [\phi'(a) - u'(a)]\, \delta(r-a)
 - [u(a)-\phi(a)]\, \delta'(r-a)\,.
\end{eqnarray}
When substituting this expression into Eq.~(\ref{Schr}), there exists
a freedom in attributing the local terms in (\ref{psi''}) to the
internal or external regions. One possible choice is to define the
singular functions as $\delta(r-a_{-})\!\equiv\!\langle
r|a_{-}\rangle\!=\!\langle a_{-}|r\rangle$ and
$\delta'(r-a_{+})=-\frac{d}{da_{+}}\,\delta(r-a_{+})\!\equiv\!-\langle
r|a_{+}\rangle'\!=\!-\langle a_{+}|'r\rangle$, where $a_{\pm}=a\pm0$,
thus considering the first as belonging to the internal subspace and
the second to the external one, respectively.

With such definitions, we can represent (\ref{Schr}) in the form
\begin{equation}\label{Htot}
\left(\begin{array}{cc} \HintN & V \\
        V^{\dagger}  & \HextD  \end{array}\right)
\left(\begin{array}{c} u \\ \phi \end{array}\right) =
k^2\left(\begin{array}{c} u \\ \phi \end{array}\right)\,,
\end{equation}
with the following entries:
\begin{subequations}\label{N}
\begin{eqnarray}
& \HintN = {\hat K} + U
 + |a_{-}\rangle\langle a_{-}|'\,,              \\
& \HextD = {\hat K} + |a_{+}\rangle'\langle a_{+}| \,,        \\ &
V=-|a_{-}\rangle\langle a_{+}|'\,, \quad
V^{\dagger}=-|a_{+}\rangle'\langle
a_{-}|\,,
\end{eqnarray}
\end{subequations}
where ${\hat K}$ stands for the kinetic energy operator. In
particular, in the coordinate representation the matrix elements of
the entries take the form
\begin{subequations} \label{CN}
\begin{eqnarray}
\langle r|\HintN|r'\rangle  &=& \left[\textstyle -\frac{d^2}{dr^2}
  + U(r)\right]\delta(r-r')%
  \nonumber\\ && -\delta(r-a_{-})\,\delta'(r'-a_{-})\,,       \\ %
\langle r|\HextD|r'\rangle &=& \textstyle
-\frac{d^2}{dr^2}\,\delta(r-r')
  \nonumber\\ && -\delta'(r-a_{+})\,\delta(r'-a_{+}) \,,       \\ %
\langle r|V|r'\rangle &=& \langle r'|{V}^{\dagger}|r\rangle =
  \delta(r-a_{-})\,\delta'(r'-a_{+})\,. %
\end{eqnarray}
\end{subequations}
The presence of the singular terms assures Hermiticity of the
operators $\Hint$ and $\Hext$ in contrast to ${\hat K}$ alone
\cite{Bloch1957}. The relations $\langle u_1|\Hint u_2\rangle\!=
\!\langle\Hint u_1|u_2\rangle$, etc., can be easily checked by means
of partial integrations. The singular operators on the boundary
provide the boundary conditions, the first one Neumann for the
internal region, the second one Dirichlet for the external region.
Equation (\ref{psi''}) implies that both BCs are interrelated. This
also leads to the adjointness of the coupling operators $V$ and
$V^{\dagger}$. The Dirichlet boundary operator is essentially
different from that of Neumann type and cannot be produced by the
boundary operators as used in \cite{Bloch1957}.

The range of operators in Eq.~(\ref{CN}) within the Hilbert space is
defined by the functions on which the singular terms vanish. This
requirement fixes the BCs at the point of separation $a$. The full
Hilbert space is a direct sum of the space spanned by the
eigenvectors of the internal problem $0<r<a$ with Neumann BC,
\begin{equation}\label{Nint}
\left(-\frac{d^2}{dr^2} + U(r)\right)\, u_n^0(r)
 = \varepsilon_n^{\textsc{(n)}} u_n^0(r),\ \ {u_n^0}'(a)=0
\end{equation}
and that spanned by the eigenvectors of the external problem $r>a$
with Dirichlet BC,
\begin{equation}\label{Dext}
-\frac{d^2}{dr^2} \phi_k^0(r) = k^2 \phi_k^0(r),\ \ \phi_k^0
(a)=0;\,\, k\geq
0\,.
\end{equation}
The type of BC is explicitly indicated by the corresponding
superscripts in Eqs.~(\ref{Htot})--(\ref{CN}). As usual, we assume
normalization conditions
\begin{subequations}
\label{norm}
\begin{eqnarray} \int_0^au_n^0(x)u_m^0(x)\ dx = \delta_{nm}\,,  \\
\int_a^{\infty}\phi_k^0(x)\phi_{k'}^0(x)\ dx = \delta(k-k')\ .
\end{eqnarray}
\end{subequations}
in the discrete and continuous spectra, respectively. The external
function
\begin{equation}
\phi_k^0(r)=\sqrt{\frac{2}{\pi}}\sin[k(r-a)]\propto
e^{-ikr}-e^{-2ika}e^{ikr}
\end{equation}
describes a wave fully reflected at the point $a$, the corresponding
$S$ function being equal to $S^{\textsc{(d)}}_0(k)=e^{-2ika}$.

One can also proceed in the opposite way and ascribe the terms with
$\delta'$ and with $\delta$ to the internal and external regions,
respectively. In this case, the entries in Eq.~(\ref{Htot}) read
\begin{subequations}
\label{D}
\begin{eqnarray}
& \HintD = {\hat K} + U - |a_{-}\rangle'\langle a_{-}|\,,        \\
& \HextN = {\hat K} - |a_{+}\rangle\langle a_{+}|'\,,            \\
&\tilde{V} = |a_{-}\rangle'\langle a_{+}|\,,  \quad  \
\tilde{V}^{\dagger} = |a_{+}\rangle\langle a_{-}|'\,.     %
\end{eqnarray}
\end{subequations}
That corresponds to the nonperturbed internal problem with Dirichlet
BC $v^0_n(a)\!=\!0$, when the external one has Neumann BC
${\varphi^0_k}'(a)\!=\!0$. The latter results in an additional shift
by $\pi/2$ of the reflection phase at the separation point $a$:
$S^{\textsc{(n)}}_0(k)=-S^{\textsc{(d)}}_0(k)=e^{-2i(ka+\pi/2)}$.

The interplay of the internal and external motions due to the
off-diagonal elements $V$, (\ref{N}\,c), distorts the outer waves
$\phi$. From the upper row in Eq.~(\ref{Htot}) we find for all
$k^2\neq\varepsilon_n$, $n$=1,2,...
\begin{equation}\label{u}
u={1\over k^2-\Hint}V\phi\ \equiv \Gint V\phi.
\end{equation}
Here $\Gint$ is the resolvent operator for the internal problem.
Then, the lower row transforms into
\begin{equation}\label{phi}
\left(k^2-\Hext - V^{\dagger}\ \Gint\ V\right)\ \phi = 0\,,
\end{equation}
or, in the position representation,
\begin{eqnarray}\label{pphi}
\textstyle (k^2 + \frac{d^2}{dr^2})\phi(r) +
\delta'(r-a_{+})[\phi(a) + \GintN (a,a) \phi'(a)]= 0\,, \nonumber \\
\textstyle (k^2 + \frac{d^2}{dr^2})\phi(r) -
\delta(r-a_{+})[\phi'(a) -{\GintD}''(a,a)\phi(a)]= 0 \nonumber
\end{eqnarray}
for the two cases considered. This yields change of the BCs of the
external part of the exact wave function to
\begin{subequations}\label{BCext}
\begin{eqnarray}
\phi(a) + \GintN(a,a)\,\phi'(a) = 0\,,       \label{BCextN} \\ %
\phi'(a) - {\GintD}''(a,a)\,\phi(a)=0\,,
\label{BCextD}    %
\end{eqnarray}
\end{subequations}
where the shorthand
\begin{equation}
{\GintD}''(a,a)\equiv {\partial^2\over \partial r\partial r'}\,
\GintD(r,r')\Big|_{r,r'\rightarrow a}\,
\end{equation}
has been used. Let us note that Eq.~(\ref{BCextN}) is the
conventional boundary condition in the Wigner-Eisenbud $R$-matrix
theory of resonance nuclear reactions \cite{Wigner1947,Lane1958}.

\section{$S-$ and $R-$functions}

In the external region $r>a$, where the potential vanishes
identically, the wave function has the form
$\phi(r)=\textrm{const}[e^{-ikr}-S(k)\,e^{ikr}]$. Therefore, the
conditions (\ref{BCext}) obtained above allow us to express the
function
\begin{equation}
S(k)\equiv e^{2i\delta(k)} =
\frac{1+ik\phi(a)/\phi'(a)}{1-ik\phi(a)/\phi'(a)}\,e^{-2ika}
\end{equation}
in terms of the internal Green's functions,
\begin{subequations}\label{SR}
\begin{eqnarray}
S(k) &=& \frac{1-ik\,\GintN(a,a)}{1+ik\,\GintN(a,a)}\,
S^{\textsc{(d)}}_0(k)\\
&=&\frac{1-(i/k)\,{\GintD}''(a,a)}{1+(i/k)\,{\GintD}''(a,a )}\,
S^{\textsc{(n)}}_0(k)\,.
\end{eqnarray}
\end{subequations}
Both expressions are exact and equivalent to each other. The only
difference is that they are written in different complete bases in
the full Hilbert space. The merits as well as limitations of each of
these representations will be discussed below.

We define further the phase shifts $\delta^{\textsc{(n,d)}}(k)$ due
to the influence of the internal region and the functions
$R^{\textsc{(n,d)}}(k)\equiv -2\tan\delta^{\textsc{(n,d)}}(k)$ by
\begin{subequations}\label{R}
\begin{eqnarray}
\label{RN} &R^{\textsc{(n)}}(k) = 2 k\,\GintN(a,a) = \textstyle 2k\,
\sum_n\frac{u^0_n(a)\,u^0_n(a)}{k^2-\varepsilon_n^{\textsc{(n)}}}\,,
\\
\label{RD}
&R^{\textsc{(d)}}(k) = \textstyle \frac{2}{k}\,{\GintD}''(a_{-},a) =
\frac{2}{k}\,\sum_n \frac{{v^0_n}'(a_{-})\,{v^0_n}'(a) }{
k^2-\varepsilon_n^{\textsc{(d)}}} \,. \mbox{\quad}
\end{eqnarray}
\end{subequations}
The spectral representations of the Green's function are used here in
the last steps. This shows that the energy levels of the
corresponding internal problems are real poles of the $R$ functions
in the complex energy plane. The formulas obtained are in close
analogy with the representations found in \cite{Pichugin2001} in a
different way. They are quite similar to those appearing in the
$R$-matrix theory \cite{Wigner1947,Lane1958}.  We should, however,
stress that functions $R$ defined in our manner differ from the
standard Wigner's $R$ function by trivial factors like $-2k$ or
$-2/k$ which we for the sake of convenience include in the
definition.

A certain caution is needed while taking the limit
$r,r'\rightarrow{a}$ in Eq.~(\ref{RD}) since the derivative of the
Green's function
\begin{eqnarray}\label{DGF}
\GintD(r,r') &=& \theta(r'-r)\chi^0_1(r)\,\chi^0_2(r') +
\theta(r-r')\chi^0_1(r')\,\chi^0_2(r) \nonumber \\
&=&\sum_n\frac{v^0_n(r)\,v^0_n(r')}{k^2-\varepsilon_n^{\textsc{(d)}}}
\end{eqnarray}
has a discontinuity when $r=r'$. The symbols $\chi^0_{1}(r)$ and
$\chi^0_{2}(r)$ stand for solutions of the internal problem with
Dirichlet BCs only at the points $r=0$ or $r=a$, respectively. It is
readily seen that the mixed second partial derivative contains a
singular contribution
\begin{equation}
\left[{\chi^0_1}'(r)\chi^0_2(r)-\chi^0_1(r){\chi^0_2}'(r)\right]\delta
(r-r')
= W\delta(r-r')\,,
\end{equation}
$W=-1$ being the Wronskian. This singularity at $r=r'=a$ must be
excluded and the second derivative must be understood as
\begin{equation}\label{lim}
{\GintD}''(a_{-},a)={\chi^0_1}'(a){\chi^0_2}'(a)\,.
\end{equation}
It immediately follows from this remark that the spectral sum
$\sum_n{v^0_n}'(r)\,{v^0_n}'(r')/(k^2-\varepsilon_n^{\textsc{(d)}})$
diverges when $r=r'$. Indeed, convergence of this sum depends on the
contributions of the very high levels, $n\rightarrow\infty$. For such
an excitation, we can neglect the potential $U(r)$ whereupon the
solution of the internal problem reduces simply to
$v^0_n(r)=\sqrt{2/a}\sin(n\pi\, r/a)$. The contribution
$\delta(r-r')$ to be dropped appears due to the factor
$\varepsilon_n^{\textsc{(d)}}=(\pi n/a)^2$ which arises in the
numerator after the double differentiation has been done. It is easy
to see, that just the contribution of this kind mainly caused
numerical problems in \cite{Pichugin2001}. For $r\neq r'$, the sum
converges, although slowly, to the finite limit (\ref{lim}) because
of oscillations \cite{convergence}.

It would be a mistake to rely upon the spectral representations in
(\ref{R}), and interpret the eigenvalues $\varepsilon_n$ as energies
of resonance states and the phases $\delta^{\textsc{(d)}}_0(k)=ka$ or
$\delta^{\textsc{(n)}}_0(k)=ka+\pi/2$ as smooth phases of the
background scattering. Indeed,  the levels $\varepsilon_n$ depend on
what sort of boundary conditions have been used. More than that, many
other forms of BC are equally permissible (see, for example, the
fundamental review \cite{Lane1958}). It can be shown that the most
general possible BC,
\begin{equation}\label{GBC}
{u^0}'(a)+\beta_{\textrm{int}}u^0(a)=0\,,\quad
{\phi^0}'(a)+\beta_{\textrm{ext}}\phi^0(a)=0\,,
\end{equation}
involve two arbitrary parameters $\beta$ (here we include formally
$\beta=\pm\infty$ for the Dirichlet BCs). Returning to
Eq.~(\ref{SR}), we see that, due to the relation
$S^{\textsc{(n)}}_0(k)=-S^{\textsc{(d)}}_0(k)$, the levels
$\varepsilon_n^{\textsc{(d)}}$ are {\it zeros} of the function
$R^{\textsc{(n)}}(k)$ and, similarly, the levels
$\varepsilon_n^{\textsc{(n)}}$ are zeros of $R^{\textsc{(d)}}(k)$. In
fact, the two phases are connected as
\begin{equation}\label{tancot}
\tan\delta^{\textsc{(d)}}(k)=
 -\cot\delta^{\textsc{(n)}}(k)=\tan[\delta^{\textsc{(n)}}(k)-\pi/2]\,.
\end{equation}
The shift $\pi/2$ just compensates the similar shift of
$\delta^{\textsc{(n)}}_0(k)$ and the total scattering phase
$\delta(k)=\delta^{\textsc{(n)}}(k)+\delta_0^{\textsc{(d)}}(k)=
\delta^{\textsc{(d)}}(k)+\delta_0^{\textsc{(n)}}(k)$ does not depend
on the type of BC used. In particular, the positions of the poles of
the function $S(k)$ in the complex $k$ plane, which are found from
the equation
\begin{equation}\label{s-pls}
1-i\tan\delta^{\textsc{(d)}}(k)=0=1-i\tan\delta^{\textsc{(n)}}(k)\,,
\end{equation}
are BC independent because of the relation (\ref{tancot}). This is in
agreement with the fact that both factors $S^{\textsc{(d,n)}}_0$ are
entire functions in $k$ plane.

The factorized form of the residues of poles in (\ref{R}) allows one
to represent the function $S(k)$ in a different but fully equivalent
form as
\begin{equation}\label{ST}
S(k) = \left(1-i\,A^T\frac{1}{k^2-\mathcal{H}}\,A\right)S_0(k) \,,
\end{equation}
where the non-Hermitian symmetric operator $\mathcal{H}$ is defined
by
\begin{equation}\label{calH}
\mathcal{H}(k)=\varepsilon -\frac{i}{2}A\,A^T\,.
\end{equation}
Here $\varepsilon$ is a diagonal matrix of the eigenvalues of the
corresponding internal problems and the column vector $A(k)$ of
coupling amplitudes, which originate from the off-diagonal element
$V$ in Eq.~(\ref{Htot}), has the components
\begin{equation}\label{Amp}
A^{\textsc{(n)}}_n(k)=\sqrt{2k}\,u^0_n(a)\,\quad\mbox{or} \quad
A^{\textsc{(d)}}_n(k)=\sqrt{\frac{2}{k}}\,{v^0_n}'(a)
\end{equation}
in the cases of Neumann or Dirichlet BC, respectively. A proof of
equivalence to the expressions (\ref{SR}) immediately comes from the
following relation between resolvents (see, for example,
\cite{Sokolov1989}):
\begin{equation}\label{GT-GR}
\frac{1}{k^2-{\cal H}}
=G_{\textrm{int}}-\frac{i}{2}\,G_{\textrm{int}}
\,A\,\frac{1}{1+\frac{i}{2}R}\,A^T\,G_{\textrm{int}}\,.
\end{equation}

It follows from Eq.~(\ref{ST}) that the poles of the $S$ function can
also be found from the secular equation
\begin{equation}\label{seq}
\det\left[z^2-{\cal H}(z)\right]=0
\end{equation}
where $z$ is a point in the complex $k$ plane. In the position
representation this is equivalent to the spectral problem with a
complex (outgoing-wave) boundary condition
\begin{subequations}\label{KP}
\begin{eqnarray}%
& \left(-\frac{d^2}{dr^2} + U(r)\right)\psi(r) = z^2\psi(r)\,, \\
& \psi'(a)-iz\psi(a)=0\,.
\end{eqnarray}
\end{subequations}
This clearly demonstrates again the independence of the choice of BCs
for disconnected internal and external motions.

Actually, the form of the operator (\ref{calH}) is provocative. One
is tempted to interpret this operator as an effective Hamiltonian
whose {\it $k$-dependent} eigenvalues define complex energies of
metastable resonance states formed by the potential $U(r)$. However,
this interpretation is, generally, wrong. To make our points clearer,
let us start with the following simple remark. Let the potential
vanish identically everywhere, $U(r)\equiv 0$, so that the $S$
function $S(k)\equiv 1$. Nevertheless, neither the operator
$\mathcal{H}$ nor the function
\begin{equation}\label{tilS}
{\tilde S}(k)=1-i\,A^T\frac{1}{k^2-\mathcal{H}}\,A
\end{equation}
are trivial in themselves. For example, in the case of Neumann BCs
matrix elements of $\mathcal{H}$ are
\begin{equation}\label{calHN0}
\mathcal{H}^{\textsc{(n)}}_{mn}=\frac{1}{a^2}\left[\left(m+\frac{1}{2}
\right)^2
\pi^2\,\delta_{mn}-ika(-1)^{(m+n)}\right]\,.
\end{equation}
Any truncated finite-size $N\times N$ matrix obtained from
Eq.~(\ref{calHN0}) gives, when substituted in Eq.~(\ref{seq}), $N$
pairs of complex roots. However, the complex poles of the resolvent
$[z^2-\mathcal{H}_{\textrm{trunc}}(z)]^{-1}$ of the truncated matrix
have nothing to do with the poles of the genuine $S$ function. The
truncation procedure is not stable when $N\rightarrow\infty$. The
difficulties become even worse in the case of Dirichlet BCs when the
imaginary part also grows,
\begin{equation}\label{calHD0}
{\cal
H}^{\textsc{(d)}}_{mn}=\frac{1}{a^2}\left[m^2\pi^2\,\delta_{mn}-
i\pi^2\frac{mn}{ka}(-1)^{(m+n)}\right]\,.
\end{equation}
In fact, all poles found go to infinity in the limit
$N\rightarrow\infty$. Indeed, the true poles, because of the identity
$\det[z^2-{\cal H}(z)]=\det(z^2-\Hint)[1+\frac{i}{2}R(z)]$, must
satisfy the equation
\begin{equation}\label{seceq}
1+\frac{i}{2}R(z)=0\,,
\end{equation}
which is equivalent to (\ref{s-pls}). For example, in the case of
Neumann BC this equation runs as $1+iz\,\GintN (a,a)=0$. In
particular, when the potential $U(r)$ vanishes identically, the
Green's function in the $z$ plane is equal to
$\GintN(a,a)=-(1/z)\tan{za}$, Eq.~(\ref{seceq}) looks as $1-i\tan
za=0$ and, therefore, has no roots in any finite domain of this
plane. This implies, in turn, that Eq.~(\ref{KP}) has no non-trivial
solutions. In fact, the function (\ref{tilS}) is simply equal to
$\tilde{S}_{\textrm{free}}(k)=e^{2ika}$ in this case and compensates
exactly the phase shift due to the fictitious reflection at the
separation point $a$.

The remark above is of quite a general nature. The radius of
separation can be chosen arbitrarily. For the sake of simplicity, we
suggest only that $a$ is larger than the finite radius of the
potential. The actual wave function satisfies at this point the
conditions
\begin{equation}\label{actBC}
\phi(a)=u(a)\,,\quad \phi'(a)=u'(a)\,.
\end{equation}
Of course, none of these quantities is known before the problem has
been solved. Any boundary condition used above generates a complete
basis in the Hilbert space, in which the actual wave functions can be
expanded. All such bases are formally equivalent. But this does not
mean that all of them are equally adequate from the physical point of
view. In particular, the more the basis eigenvectors and their
derivatives differ at the point $a$ from the real values
(\ref{actBC}) the more slowly the corresponding expansion converges
near this point.

A forced BC creates a false reflection at this point, which is
described by the factor $S_0(k)$, whose phase should be fully
compensated by the similar part of the total phase of the function
${\tilde S}(k)$. Both factors separately depend on $a$ although the
complete function $S(k)$ does not. The main role in this compensation
is played by the matrix elements ${\cal H}_{mn}$ with large $m,n$
when the influence of the potential $U$ becomes negligible, and we
return to the situation described in the previous paragraph.

However, the choice of the separation radius and BC influences the
positions of the poles (as well as the residues) of the function $R$
and, consequently, the explicit form of the matrix
$\mathcal{H}_{mn}$. This influence is stronger the less adequate the
choice made of BC and $a$. In general, the parameters of the $R$
function can carry rather poor information about the actual complex
poles. For this reason, diagonalization of a truncated matrix
$\mathcal{H}_{mn}$ (which is necessary in any numerical computation)
can lead in the case of a poor choice to strong disagreement with the
characteristics of the actual poles of the function $S$. The explicit
dependence of the matrix elements $\mathcal{H}_{mn}$ on the wave
number $k$ causes additional problems (see next section). To extract
the physically relevant effective Hamiltonian from the formal
operator $\mathcal{H}$, additional physical considerations must be
engaged. For example, one may expect from the physical point of view
that the most relevant choice of the separation point $a$ would be a
distance matching an outer potential barrier which is strong enough
to make immediate reflection at this point quite probable.

To explore in more detail the questions briefly discussed above, we
will apply in the next section the formal technique sketched here to
the problem of scattering by a finite periodic set of $\delta$
barriers, which can be solved exactly.

\section{Open Kronig-Penney model}

\subsection{Exact solution}

We will consider below $s$-wave scattering by a periodically disposed
chain of $L$ $\delta$ barriers,
\begin{equation}\label{U}
U(r)=\sum_{l=1}^{L}\kappa_l\delta(r-l)\,.
\end{equation}
To ensure formation of long-lived resonance states, at least some of
the strength constants $\kappa_l$ should be positive. The distance
$r$ is measured in units of the size of the well formed by two
neighboring barriers. Below, we derive the effective Hamiltonian
starting directly from the Schr\"odinger equation.

Because of the local character of the barriers, it is most convenient
to start from, instead of the Schr\"odinger equation, the equivalent
integral equation
\begin{eqnarray}\label{Inteq}
\chi(r) &=&\sin(kr)+\int_0^{\infty}dr' G^0_{+}(r,r')\,U(r')\,\chi(r')
\nonumber\\
&=& \sin(kr)+\sum_{l=1}^{L}\kappa_l G^0_{+}(r,l)\,\chi(l)\,.
\end{eqnarray}
Here the symbol $G^0_{+}(r,r')=-(1/k)\,g^0(r,r')$ with
\begin{equation}\nonumber
g^0(r,r')=\theta(r'-r)\sin(kr)e^{ikr'}+\theta(r-r')\sin(kr')e^{ikr}
\end{equation}
stands for the Green's function of the free radial motion which has
outgoing-wave asymptotic. From the second line in (\ref{Inteq}) we
find immediately $S(k)=1-iT(k)$ where the scattering amplitude is
given by
\begin{equation}\label{T}
T(k)=\frac{2}{k}\sum_{l=1}^{L}\kappa_l\sin(kl)\chi(l) =
\frac{2}{k}\,s^{\textsc{t}}\kappa\,\chi\,.
\end{equation}
In the second equality we have used matrix notation, with $s$ and
$\chi$ being $L$-dimensional vectors with the components $s_l\equiv
s_l(k)\equiv \sin(kl)$ and $\chi_l\equiv\chi(l)$ ($l$=1,2,...,$L$),
respectively, when
$\kappa=\textrm{diag}\{\kappa_1,\kappa_2,...,\kappa_{\textsc{l}}\}$.

According to Eq.~(\ref{Inteq}), the $L$-dimensional vector $\chi$
satisfies the equation $\chi = s+G^0\kappa\chi$, where
$G^0=-(1/k)g^0$ and $g^0$ is a symmetric non-Hermitian matrix with
the matrix elements
\begin{equation}\label{eqchi}
g^0_{ll'}=\left\{\begin{array}{c} \sin(kl)\,e^{ikl'} \quad\mbox{if}
\quad l\leq l'\,,  \\ \sin(kl')\,e^{ikl} \quad\mbox{if}\quad l>l'\,.
\end{array}\right.
\end{equation}
Thus, we obtain finally
\begin{subequations}\label{exT}
\begin{eqnarray}
S(k)&=&e^{2i\delta (k)} = 1-is^{\textsc{t}}\mathcal{G}(k)s =
\frac{1-\frac{i}{2}K(k)}{1+\frac{i}{2}K(k)}\,,\mbox{\quad}\\
K(k)&=&-2\tan\delta(k) = s^{\textsc{t}}G(k)s\,,
\end{eqnarray}
\end{subequations}
where we have used the factorized form $\Imm{g^0}=ss^{\textsc{t}}$ of
the anti-Hermitian part of the matrix $g^0$ to pass from the first to
the second equality. The $L\times L$ matrix propagators
\begin{equation}\label{props}
\mathcal{G}(k)=\frac{2}{k\lambda+g^0}\quad \mbox{and} \quad
G(k)=\frac{2}{k\lambda+\Ree{g^0}}
\end{equation}
are connected with one another by a relation similar to
(\ref{GT-GR}). The diagonal matrix $\lambda=\kappa^{-1}$
characterizes the penetrabilities of the barriers \cite{barrier}. The
poles of the $S$ function are defined by the equation
\begin{equation}\label{Sroots}
\textstyle \det[z\lambda+g^0(z)]=0=1+\frac{i}{2}\,K(z)\,.
\end{equation}

All the matrix elements (\ref{eqchi}) are entire functions in the
complex $z$ plane. The same is valid for the determinant in
(\ref{Sroots}). Therefore, one can show that this equation has an
infinite number of isolated complex roots. Further, its roots because
of the relation $[g^0_{ll'}(z)]^*=-g^0_{ll'}(-z^*)$ come in pairs
$z_n$ and $z_{-n}\equiv -z^*_n$, $n$=1,2,..., symmetrically with
respect to the imaginary axis, or lie on the latter, $z_q=iy_q$. All
poles of the first type are situated in the lower part of the complex
plane. Poles on the positive half of the imaginary axis, $y_q>0$,
correspond to bound states and can appear only if some number of the
constants $\kappa_l$ are negative. Those which are situated on the
negative part, $y_q<0$, correspond to the so-called virtual levels.
The total number of purely imaginary poles is finite for any finite
$L$.

After all poles have been found, the function $S(k)$ can be presented
in the form
\begin{equation}\label{factS}
S(k)=\prod_{n=1}^{\infty}\frac{(k+z_n)(k-z^*_n)}{(k-z_n)(k+z^*_n)}\,
\prod_{q} \frac{k+iy_q}{k-iy_q}\,
\end{equation}
where we took into account that $S(k\!=\!\infty)=1$, since $\delta$
barriers become transparent for a particle with asymptotically large
energy. These expressions are in agreement with the general theory
\cite{Hu1948,Nussenzveig1972}, which is valid for any potential with
a finite radius.

Each factor in (\ref{factS}) is singly unitary. In particular, for a
pair $z_{\pm n}$ of conjugate roots we have
\begin{eqnarray}\label{Sn}
S_n(k) &\equiv& \frac{(k+z_n)(k-z^*_n)}{(k-z_n)(k+z^*_n)}
\nonumber\\
&=&\frac{k^2-|z_n|^2+2ik\,\Imm z_n}{k^2-|z_n|^2-2ik\,\Imm z_n}\equiv
e^{2i\delta_n(k)}\,.
\end{eqnarray}
Since $\Imm{z_n}<0$, the phase $\delta_n(k)$ increases when the
energy $E=k^2$ grows and passes the value $\pi/2$ at the point
$E=|z_n|^2\equiv E_n$. The typical energy interval $\Delta E$ of the
main gain of the phase is estimated by the quantity $2k|\Imm{z_n}|$.
If this interval is small enough and the latter quantity does not
vary appreciably within it, the total gain is close to $\pi$ and the
factor $S_n$ receives the standard Breit-Wigner resonance form
\begin{equation}\label{resSn}
S^{(res)}_n(k)=\frac{E-E_n-\frac{i}{2}\Gamma_n}{E-E_n+\frac{i}{2}
\Gamma_n} \equiv \frac{E-{\cal E}^*_n}{E-{\cal E}_n}\,,
\end{equation}
with the energy $E_n\equiv\Ree{\mathcal{E}_n}$ and width
$\Gamma_n\equiv -2\Imm{\mathcal{E}_n}$ of the resonance defined as
follows
\begin{equation}\label{EnWth}
E_n=|z_n|^2\,,\quad \Gamma_n=4|\Ree{z_n}\Imm{z_n}|\,.
\end{equation}
The $k$-dependence is neglected in the definition of the width and
substitution $k=\sqrt{E}\approx|\Ree{z_n}|$ has been made. Such a
substitution is well justified when the scattering energy is large
enough, but becomes improper near $E=0$. Due to the ``threshold''
$\sqrt{E}$ dependence of the widths, some specific behavior takes
place when in the proximity of this point there exists a number of
bound and decaying states
\cite{Sokolov1992,Persson1996ii,Zelevinsky2002}.

We conclude that the scattering amplitude $T(k)=i[S(k)-1]$ is a
meromorphic function in the $z$ plane and can therefore also be
written down as an infinite sum of the pole contributions
\begin{equation}\label{prepT}
T(k) = \sum_{n=-\infty}^{\infty}
\frac{\varrho_n}{k-z_n}+\sum_q\frac{\varrho_q}{k-iy_q}\,,
\end{equation}
where the residues $\varrho_{n,q}$ can easily be found from
Eq.~(\ref{factS}).

Another representation of the scattering amplitude sometimes
considered in the literature ensues from diagonalization of the
(finite in our case) matrix ${\cal G}(k)$ at a given fixed real $k$.
Since this matrix is complex symmetric, such a diagonalization is
performed by a complex orthogonal transformation defined together
with the {\it complex} eigenvalue matrix
$\Lambda(k)=\textrm{diag}\{\Lambda_1(k),\Lambda_2(k),...,\Lambda_L(k)\}$
from the equation
\begin{equation}
[k\lambda+g^0(k)]\Psi(k)=\Psi(k)\Lambda(k)\,.
\end{equation}
The scattering amplitude reduces then to a finite sum
\begin{equation}\label{fprepT}
T(k)=2\sum_{l=1}^L\frac{[s^{\textsc{t}}(k)\Psi^{(l)}(k)]^2}
{\Lambda_l(k)}\,,
\end{equation}
where $\Psi^{(l)}(k)$ is the $l$th eigenvector. However, such a
representation is of limited practical use inasmuch as the terms of
the sum are, generally, extremely complicated. There is no simple
interpretation of a given term and all of them can give comparable
contributions. The compatibility of the two representations
(\ref{prepT}) and (\ref{fprepT}) is very indirect. A particular term
of the second one cannot be uniquely continued from the real axis at
an arbitrary point $z$ in the complex plane because both
$\Psi^{(l)}(z)$ and $\Lambda_l(z)$ are multivalued functions in this
plane. The same is, of course, also true for the opposite direction.
Therefore, even if a root $z^l_j$ of the equation $\Lambda_l(z)=0$ is
found and near this point the pole approximation
\begin{equation}\label{fxpnt}
T^l_j(z)\approx \frac{2\,[s^{\textsc{t}}(z^l_j)\Psi^{(l)}(z^l_j)]^2/
\Lambda'_l(z^l_j)}{z-z^l_j}
\end{equation}
is valid, this expression cannot, generally speaking, be continued on
the real axis by simply substituting $z\rightarrow k$, since the
power expansion
$\Lambda_l(z)=\Lambda_l(z^l_j)+\Lambda'_l(z^l_j)(z-z^l_j)+...$ has a
finite radius of convergence. Further, for any fixed $l$ a set of
roots exists depending on the branch of the function $\Lambda_l(z)$
considered. There is no one-to-one correspondence between the set of
roots $z^l_j$ and the manifold of actual poles $z_{n,q}$ that appears
in the expansion (\ref{prepT}). Many of the roots $z^l_j$ are false
and their contributions must finally cancel out. Therefore,
diagonalization of the propagator with non-trivial energy dependence
is not as a rule useful; rather it can lead to misleading
conclusions.

We compare below the closed expressions Eq.~(\ref{exT}) found with
the formalism described in the previous section. In correspondence
with our remark at the end of Sec.~III, we fix the separation point
$a$ by superposing it on the position of the outer barrier, $a=L$.
The appearance of an additional $\delta$ function changes the
corresponding boundary conditions. In particular, if the
$\delta$-terms are ascribed to the internal region, Neumann BC
(\ref{Nint}) for the internal problem is replaced by
\begin{equation}\label{cNint}
\kappa_{\textsc{l}}u^0(a)+{u^0}'(a) = 0 =
u^0(a)+\lambda_{\textsc{l}}{u^0}'(a)\,,
\end{equation}
whereas attributing such terms to the external domain  yields the
following change of the BC of the external problem:
\begin{equation}\label{cNext}
\kappa_{\textsc{l}}\phi^0(a)-{\phi^0}'(a) = 0 =
\phi^0(a)-\lambda_{\textsc{l}}{\phi^0}'(a)\,.
\end{equation}

We stress that neither of these conditions coincides with the
boundary conditions
\begin{equation}\label{exactBC}
\begin{array}{l}
\chi(a_{-})=\chi(a_{+})\equiv\chi(a)\,,\\[1ex]
\chi'(a_{+})-\chi'(a_{-})=\kappa_{\textsc{l}}\chi(a)\,
\end{array}
\end{equation}
satisfied by the exact wave function $\chi(r)$.

In the case (\ref{cNext}), the immediate reflection at the point $a$
is described instead of $S^{\textsc{(n)}}_0$ by
\begin{equation}\label{vlev}
S_0(k)=S^{\textsc{(n)}}_0\frac{k-i\kappa_{\textsc{l}}}
{k+i\kappa_{\textsc{l}}} = \frac{1+i\lambda_{\textsc{l}}k}
{1-i\lambda_{\textsc{l}}k}\,S_0^{\textsc{(d)}}\,.
\end{equation}
(Recall that $S_0^{\textsc{(d)}}(k)=e^{-2ika}\equiv e^{-2ikL}$.) The
additional factor is due to the influence of the barrier outside the
radius of separation $a$. This factor has a pole
$k=-i\kappa_{\textsc{l}}$ on the negative part of the imaginary axis,
which formally corresponds to a virtual Wigner level. Actually, such
a pole does not exist in the exact solution and disappears due to
cancellation with the contribution of the internal region [see
Eq.~(\ref{DlamS}) below].

\subsection{Internal problem with Neumann BC}

The exact $S$ function in the case of the BC (\ref{cNint}) reads as
$S(k)=\tilde{S}(k)S^{\textsc{(d)}}_0(k)$, where
\begin{equation}\label{lamS}
\tilde{S}(k) = \frac{1-\frac{i}{2}R}{1+\frac{i}{2}R} =
1-i\tilde{T}(k)\,,
\end{equation}
with
\begin{eqnarray}\label{lamR}
R(k)&=&\tilde{A}^{T}\frac{1}{k^2-\HintN}\tilde{A} =2k\,\GintN(a,a)
\\
\nonumber &=& \textstyle
2k\lambda_{\textsc{l}}^2\sum_{n=1}^{\infty}\frac{{u^0_n}'(a)\,{u^0_n}'
(a)}{k^2-\varepsilon_n^{\textsc{(n)}}}
\end{eqnarray}
and
\begin{equation}\label{lamT}
\tilde{T}(k)=
\tilde{A}^{T}\frac{1}{k^2-\mathcal{H}^{\textsc{(n)}}}\,\tilde{A}\,.
\end{equation}
The coupling amplitudes are equal to
\begin{equation}\label{lamA}
\tilde{A}_n(k)=\sqrt{2k}\,u^0_n(a)=
-\lambda_{\textsc{l}}\sqrt{2k}\,{u^0_n}'(a)\,.
\end{equation}
Finally, the matrix elements of the operator
$\mathcal{H}^{\textsc{(n)}}$  [see Eq.~(\ref{calH})]  appear as
\begin{eqnarray}\label{lamcalH}
\mathcal{H}^{\textsc{(n)}}_{mn} &=&
\varepsilon_m^{\textsc{(n)}}\delta_{mn} - ik\,u^0_m(a)\,u^0_n(a)
\nonumber\\ &=& \varepsilon_m^{\textsc{(n)}}\delta_{mn} -
i\lambda_{\textsc{l}}^2k\,{u^0_m}'(a)\,{u^0_n}'(a)\,,
\end{eqnarray}
with the levels $\varepsilon_n^{\textsc{(n)}}$ being the eigenvalues
of the internal problem with the BC (\ref{cNint}). For the sake of
simplicity, we use the same superscript N as before.

In spite of the seemingly similar general structure of the
expressions (\ref{exT}), (\ref{props}) on one hand side and
(\ref{lamS})--(\ref{lamT}) on the other, they are, in essence, quite
different. The most important distinction shows itself in the
dimension of the vectors and matrices which is {\it finite} and
coincides with the number of barriers in the first case and {\it
infinite} in the second. In addition, the factor $S_0(k)$ of the
immediate reflection does not appear explicitly in Eq.~(\ref{exT}).
We will analyze below a couple of simple special cases before drawing
general conclusions.

\subsection{One $\delta$ barrier}

In this case we find immediately from Eqs.~(\ref{exT}) ($a=L=1$ and
we drop the subscript $L$ in the strength of the barrier)
\begin{subequations}\label{1S}
\begin{eqnarray}
S(k) &=& \frac{\sin k\, e^{-ik}+\lambda\, k}{\sin k\, e^{ik}+\lambda\,
k}\,, \\
K(k) &=& \frac{2\sin^2k}{\sin k\cos k+\lambda k}\,,
\end{eqnarray}
\end{subequations}
whereas
\begin{subequations}\label{1R}
\begin{eqnarray}
{\tilde S}(k) &=& \frac{\sin k+\lambda k\,e^{ik}}{\sin k+\lambda
ke^{-ik}}\,, \\
R(k) &=& -2\lambda\frac{k\sin k}{\sin k+\lambda\,k\cos k}\,.
\end{eqnarray}
\end{subequations}
In the Eqs.~(\ref{1R}) the phase $\delta^{\textsc{(d)}}(k)=-k$ of the
immediate reflection at the point $a$ is extracted. The two functions
$K(k)$ and $R(k)$ are related to each other by
\begin{equation}\nonumber
K(k) = \frac{2\tan k+R(k)}{1-\frac{1}{2}\tan k\,R(k)}\,,
\end{equation}

The positions of the poles of the function $S$ coincide with the
complex roots of the equation
\begin{equation}\label{deq}
e^{2iz}+2\,i\lambda\,z-1=0\,.
\end{equation}
In the right half of the $z$ plane they can be searched for in the
form $z_n=n\pi+\zeta_n$, where any $\zeta_n$ is restricted to the
strip $|\Ree\zeta_n|\leq\pi/2$ and satisfies the equation
\begin{equation}\label{zeta}
\zeta_n+\frac{i}{2}\ln\left[1-2i\lambda(n\pi+\zeta_n)\right]=0\,.
\end{equation}
There also exists the trivial root $z=0$ but this root fully cancels
out. Even for not very large $n$ one can omit $\zeta_n$ in the
logarithmic term. This yields the approximate solution
\begin{equation}\label{appzeta}
\zeta_n \cong -\frac{i}{2}\ln\left(1-2i\lambda n\pi\right)
\end{equation}
which is valid with good accuracy for almost all poles, being
asymptotically exact when $n\rightarrow\infty$.

However, by no means all of the complex roots $z_n$ correspond to
resonances. For a pole to correspond to a long-lived resonance state,
the following two additional conditions also must be satisfied: i)
$\Ree{z_n}$ should be a real pole of the function $K(k)$ or at least
should be close to such a pole; ii) the corresponding residue should
be small enough for the scattering phase $\delta(k)$ to increase near
this point by $\pi$ in an interval $\Delta k\ll 1$. The validity of
these conditions depends on the value of the parameter $\lambda$.
Indeed, the poles of $K(k)$ are found from the equation
\begin{equation}\label{polR1}
\sin k\cos k+\lambda k=0\,.
\end{equation}
There exists only a {\it finite number} of real roots $\pm k_j$,
$j$=1,2,...,$j_{max}$ of this equation. They satisfy the requirement
$2\lambda k_j<1$ or, equivalently, $k_j<\kappa/2$. (The trivial
solution $k$=0 should be dismissed.) All other roots lie in the
complex plane and the phase does not reach its maximal value although
can change rather fast if a pole of $K$ is still close to the real
axis. It is clear that they move away from this axis when $|z|$
grows. In particular, no real solutions exist if
$\lambda>\lambda_0\approx 0.2$ $(\kappa<\kappa_0\approx 5)$. The
scattering phase $\delta(k)<\pi/2$ in this case and smoothly depends
on $k$. The barrier is too weak to form a long-lived resonance state.
When $\kappa|\sin k/k|<1$ the scattering phase can be calculated as
$\delta(k)\approx-2\kappa\sin^2k/k+\cdots\,$ in the framework of
perturbation theory.

There are two kinds of real roots of Eq.~(\ref{polR1}) when $\lambda$
is appreciably less than $\lambda_0$. The most interesting case
$\lambda k_j\ll 1$, although $j\gg 1$, can easily be considered
analytically. The first set consists of the roots $k_n$ which are
close to the points $n\pi$: $k_n\approx(1-\lambda+\lambda^2)n\pi$,
($n\pi<1/\lambda\,$). Near such a pole the $K$ function manifests
typical resonance behavior
\begin{equation}\label{Rn(k)}
K(k)\approx \frac{2(\lambda n\pi)^2}{k-k_n}\,.
\end{equation}
Strictly speaking, a similar contribution of the symmetric root
$k_{-n}=-k_n$, which corresponds to the same energy
$\varepsilon_n=k^2_n$, should be added, so we arrive near the $n$th
resonance at
\begin{equation}\label{Rn(E)}
K_n(E)= \frac{(2\lambda)^2 (n\pi)^3}{E-\varepsilon_n}\,.
\end{equation}
Each neighboring pair of resonance roots is separated by a root of
the second set: $k_m\approx[1+(\lambda
m\pi)^{-1}](m+\frac{1}{2})\pi$, in a vicinity of which
\begin{equation}\label{Rm(k)}
K(k)\approx \frac{2}{k-k_m}\,.
\end{equation}
The residues are large in this case and such terms contribute into
the smooth background part of the total scattering phase. Indeed, the
(dimensionless) time delay
$\tau(k)$=$d\delta(k)/dk$=$\tau_w/|\tau_0|$, which measures the
Wigner time delay $\tau_w$=$2d\delta/dE$ in units of the (negative)
time delay $\tau_0$=$-2a/(dE/dk)$ because of the immediate reflection
at the point $a$\,(=1), is large near the points $k_n$ but small when
$k\!\approx\!k_m$. Poles of the two different kinds alternate and
resonant and smooth contributions are mixed. Figure 1 clearly
demonstrates all the features described above.
\begin{figure}[t] \label{fig1}
\unitlength 1cm
\begin{picture}(8,12.5)
\epsfxsize 7.5cm\put(0,-0.2){\epsfbox{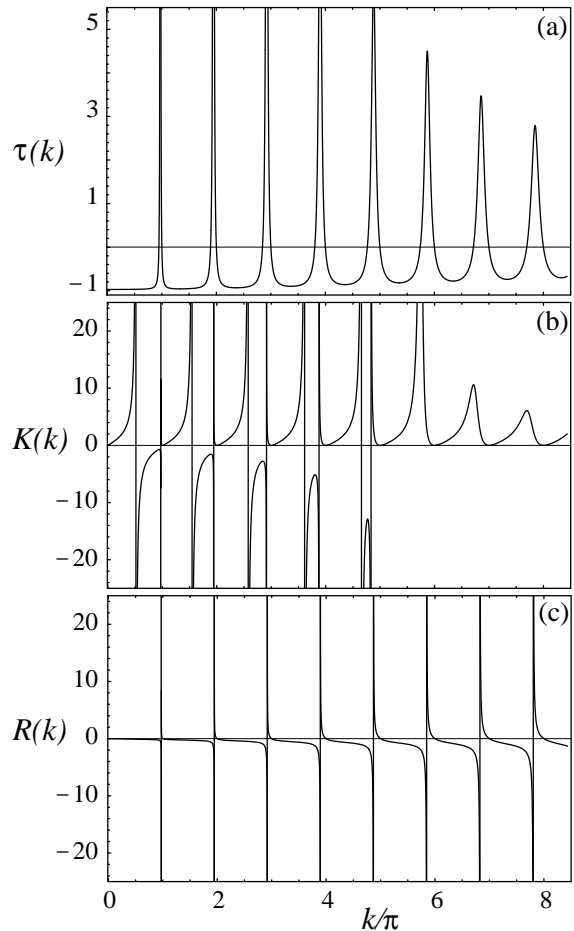} }
\end{picture}
\caption{One $\delta$ barrier model for $\lambda$=1/35. (a) The time
delay $\tau(k)$=$d\delta(k)/dk$, (b) $K$ function and (c) $R$
function. Only a finite number
$n_{\textrm{res}}\!\approx\![1/2\pi\lambda]\,(=5)$ of poles located
near the points $\pi n$, $n$=1,...,$n_{\textrm{res}}$, correspond to
long-lived resonance states; the others refer to the smooth phase of
reflection at the point $a$(=1).}
\end{figure}

Let us on the other hand consider the properties of the function
$R(k)$  [see Eq.~(\ref{1R})]. Its poles are the roots of the equation
\begin{equation}\label{poltilR1}
\sin k+\lambda\,k\cos k=0
\end{equation}
instead of Eq.~(\ref{polR1}). Contrary to the latter, {\it all} roots
$\pm\tilde{k}_n$ of (\ref{poltilR1}) are real. Those of them, that
satisfy the inequality $\lambda \tilde{k}_n\ll 1$ and, therefore,
have numbers $n<n_{max}$, are equal to
$\tilde{k}_n\approx(1-\lambda)n\pi$ again and correspond to
resonances. The difference from the similar roots of
Eq.~(\ref{polR1}) appears only at higher order corrections in the
parameter $\lambda$. For larger $n$, when $\lambda\tilde{k}_n\gg 1$,
an infinite number of roots
$\tilde{k}_n\approx\pi(n+\frac{1}{2})[1+\lambda^{-1}(n\pi)^{-2}]$
exists, giving a smooth contribution, which combines with the phase
$\delta^{\textsc{(d)}}_0(k)=-k$ and almost compensates it at large
$k$  [see Fig.~1(c)]. Indeed, the $\delta$ barrier becomes almost
transparent when the collision energy is large and the total
scattering phase can be calculated by applying perturbation theory.

The meromorphic character of the function $R(k)$ allows us to
represent this function in the form of the pole expansion
\begin{eqnarray}\label{polexp}
R(k) &=& \sum_{n=-\infty}^{\infty}
\frac{2}{1+\lambda+(\lambda\tilde{k}_n)^2}\,
\frac{(\lambda\tilde{k}_n)^2}{k-\tilde{k}_n}  \nonumber\\  &=&
2k\sum_{n=1}^{\infty}\frac{2}{1+\lambda+(\lambda\tilde{k}_n)^2}\,
\frac{(\lambda\tilde{k}_n)^2}{k^2-\tilde{\varepsilon}_n}\,.
\end{eqnarray}
The symmetry connection ${\tilde k}_{-n}=-{\tilde k}_n$ has been
taken into account in the second equality.

Returning to the poles of $S(k)$, we can expand the expression
(\ref{appzeta}) in the resonant region $2\lambda n\pi\ll 1$ with
respect to this parameter to calculate the poles
\begin{equation}\label{respls}
z_n\approx (1-\lambda+\lambda^2)\,n\pi-i(\lambda n\pi)^2
\end{equation}
and the complex energies of the resonance states
\begin{equation}\label{cmpE}
\mathcal{E}_n\approx \textstyle
(1-2\lambda+3\lambda^2)(n\pi)^2-\frac{i}{2}
[4\lambda^2(n\pi)^3]\,.
\end{equation}
Note that the widths $\Gamma_n=4\lambda^2(n\pi)^3$ of the resonances
appear only in the second order in the penetrability parameter
$\lambda$ when the shifts of their energies are of the first order of
magnitude. The resonances are well isolated since the ratio
$\Gamma_n/(E_{n+1}-E_n)=2\pi (\lambda n)^2\ll 1$. Finally, the remote
poles with $n\gg 1/2\pi\lambda$ are given by
\begin{equation}\label{aspls}
\textstyle %
z_n\approx (n-\frac{1}{4})\pi-\frac{i}{2}\ln(2\lambda n\pi)^2\,.
\end{equation}

Now we compare our findings with the results
(\ref{lamS})--(\ref{lamT}) of the general formalism. The normalized
solutions of the internal problem with BC (\ref{cNint}) are readily
found to be
\begin{equation}\label{cNsol}
u^0_n(r)=\sqrt{\frac{2}{1+\lambda\cos^2k_n}}\sin(k_nr)
\end{equation}
where $k_n$ are the roots of Eq.~(\ref{poltilR1}) which follows
directly from the BC (\ref{cNint}) this time (we omit the tilde and
note that only positive roots $k_n$, $n$=1,2,..., are to be kept).
This fact is quite satisfactory and demonstrates the physical
relevance of the choice of BC made. Indeed, for such a BC
$u^0_n(a)\sim\lambda$ in accordance with the exact boundary condition
(\ref{exactBC}). The Dirichlet BC $u^0_n(a)=0$ would be deficient in
this sense. With Eq.~(\ref{cNsol}) taken into account, we find from
Eq.~(\ref{lamR})
\begin{equation}\label{lamR1}
R(k) = 2k \sum_{n=1}^{\infty}
\frac{2\cos^2k_n}{1+\lambda\cos^2k_n}\,\frac{(\lambda k_n)^2}
{k^2-\varepsilon_n}\,.
\end{equation}
Equivalence to Eq.~(\ref{polexp}) is seen from the relation
$\cos^2k_n=[1+(\lambda k_n)^2]^{-1}$, which follows directly from the
secular equation (\ref{poltilR1}).

\begin{figure} \label{fig2}
\unitlength 1cm
\begin{picture}(8,5)
\epsfxsize 7.5cm \put(0.35,-0.2){\epsfbox{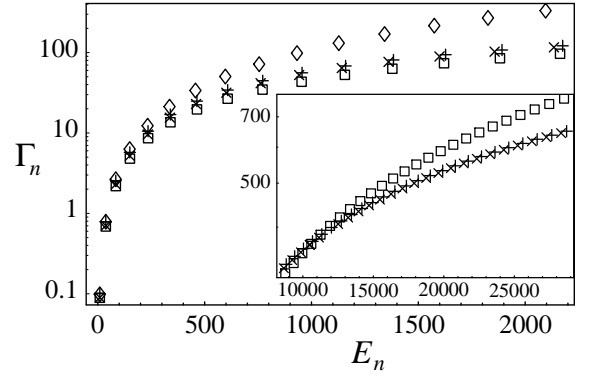} }
\end{picture}
\caption{Widths (in log scale) versus energies for $\lambda$=1/35.
Exact poles [from Eq.~(\ref{zeta}), $\square$] and approximation
[Eq.~(\ref{cmpE}), $\lozenge$], compared to the ``diagonal''
approximation $\mathcal{H}_{nn}(k\!=\!k_n)$ with Neumann [from
Eq.~(\ref{lamcalH1}), $\times$] or Dirichlet BC [from Eq.
(\ref{lamcalHD}), $+$]. Inset shows the asymptotic region of remote
poles, where the diagonal approximation breaks down.}
\end{figure}
Matrix elements of the operator $\mathcal{H}$ (\ref{lamcalH}) are in
our case
\begin{equation}\label{lamcalH1}
\mathcal{H}^{\textsc{(n)}}_{mn} = k^2_n\delta_{mn}-
i\,2k\frac{\lambda k_m}{\sqrt{1+\lambda+(\lambda k_m)^2}}\,
\frac{\lambda k_n}{\sqrt{1+\lambda+(\lambda k_n)^2}}\,.
\end{equation}
In the resonance domain, $\lambda k_n\lesssim 1$, the diagonal matrix
element $\mathcal{H}^{\textsc{(n)}}_{nn}(k\!=\!k_n)$ approximates
well the complex energy $\mathcal{E}_n$ of the resonance which lies
near the scattering energy $E\approx k^2_n$. In this region
off-diagonal elements influence only corrections of higher orders. At
the same time, the diagonal approximation becomes insufficient for
the remote poles, breaking down to reproduce exact asymptotic $n\ln
n$ behavior of the imaginary parts. Figure 2 illustrates the
consideration.

We therefore conclude that the extraction of the phase
$\delta^{\textsc{(d)}}_0(k)$ of the immediate reflection and
utilization of the $R-$function is adequate only in the resonance
region where such a reflection is probable. Beyond this region the
internal and immediate reflection phase shifts almost compensate each
other so that the function $K(k)$ proves to be a more relevant tool.

\subsection{Resonance domain}

Below, we restrict our attention to the resonance region well above
the energy $E\!=\!k^2\!=\!0$. To simplify the investigation further,
we suggest assuming the strength constants to be the same
$\lambda_l=\lambda_1$, $l\!=\!2,3,...,L\!-\!1$, for all inner
barriers. In accordance with the results of the simple consideration
presented at the end of Sec.~IV C, segregation of the smooth phase of
the immediate reflection meets expectations based on physical
intuition. To perform this segregation, we write first the scattering
amplitude $T(k)=s^{\textsc{t}}{\cal G}(k)s$ as a block product
\begin{eqnarray}\label{blcalG}
T(k) &=& ({\bf s}^{\textsc{t}},\,s_{\textsc{l}})
\left(\begin{array}{cc} {\cal {\hat G}} & \mbox{\boldmath ${\cal
F}$}
\\ \mbox{\boldmath ${\cal F}$}^T  & {\cal G}_{\textsc{l}}
\end{array}\right) \left(\begin{array}{c} {\bf s} \\ s_{\textsc{l}}
\end{array}\right) \nonumber \\
&=& \Bigl[ (\lambda_{\textsc{l}}k - s_{\textsc{l}}
e^{ika}) {\bf s}^{\textsc{t}}\frac{1}{\lambda_1 k {\hat I}+{\hat
g}^0}\,{\bf s}+s^2_{\textsc{n}}\Bigr] \mathcal{G}_{\textsc{l}},
\end{eqnarray}
where the symbols ${\cal {\hat G}}$ and $\mbox{\boldmath ${\cal
F}$}$, etc., stand for $(L-1)\times(L-1)$ submatrices and
$(L-1)$-dimensional vectors, respectively, and ${\hat g}^0$ is the
similar upper left block of $g^0$. To pass from the first to the
second line, the relations
\begin{equation}
\nonumber %
{\bf s}^{\textsc{t}}{\cal {\hat G}}{\bf s} =
-\left(\lambda_{\textsc{l}} k e^{-ika}+s_{\textsc{l}}\right)\, {\bf
s}^{\textsc{t}}\mbox{\boldmath${\cal F}$}
\end{equation}
and
\begin{equation}
\nonumber %
{\bf s}^{\textsc{t}}\mbox{\boldmath ${\cal F}$} =
\mbox{\boldmath${\cal F}$}^{\textsc{t}}{\bf s} = -e^{ika}{\bf
s}^{\textsc{t}}\frac{1}{\lambda_1 k {\hat I}+{\hat g}^0}\,{\bf s}
\end{equation}
have been used which follow, together with the expression
\begin{equation}\label{calGN}
\nonumber %
\mathcal{G}_{\textsc{l}} = 2\,\bigl[\lambda_{\textsc{l}} k
e^{-ika}+s_{\textsc{l}} -e^{ika} {\bf s}^{\textsc{t}}
\frac{1}{\lambda_1 k {\hat I}+{\hat g}^0}\,{\bf
s}\bigr]^{-1}\,e^{-ika}\,,
\end{equation}
from the equation $(k\lambda +g^0)\,\mathcal{G}(k)=2$. We notice now
that the function
\begin{equation}\label{Tinf}
T_{\infty}(k)={\bf s}^{\textsc{t}}\frac{2}{\lambda_1 k {\hat
I}+{\hat
g}^0}\,{\bf s}
\end{equation}
coincides with the amplitude of scattering on the potential (\ref{U})
with the last barrier being removed: $\kappa_{\textsc{l}}=0$ or, in
other words, $\lambda_{\textsc{l}}=\infty$. Finally,  we obtain after
simple transformations $S(k)=\tilde{S}(k)\,e^{-2ikL}$, where
\begin{eqnarray}\label{explamS}
\tilde{S}(k) &=&
\frac{e^{-2ikL}-(1+2i\lambda_{\textsc{l}}k)S_{\infty}(k) }{
(1-2i\lambda_{\textsc{l}}k)e^{-2ikL}-S_{\infty}(k) } \nonumber \\
&=&\frac{\sin(\delta_{\infty}+ka)+\lambda_{\textsc{l}}k
e^{i(\delta_{\infty}+ka)} }{
\sin(\delta_{\infty}+ka)+\lambda_{\textsc{l}}k
e^{-i(\delta_{\infty}+ka)}}\,. \mbox{\quad}
\end{eqnarray}

The poles of the function $\tilde{S}(k)$ [as well as of $S(k)$] in
the complex $k$ plane are given by zeros of the denominator. It is
convenient to introduce in parallel with this function a sequence of
functions $\tilde{S}^{(l)}(k)\equiv S^{(l)}(k)\,e^{2i l k}$ which
describe the scattering on chains of $l$ barriers with the
penetrability constant $\lambda_1$ and the BC fixed at the radius
$a=l$. In particular, $\tilde{S}^{(L)}(k)\equiv\tilde{S}(k)$,
$\tilde{S}^{(L-1)}(k)\equiv S_{\infty}(k)\,e^{2i(L-1)\, k}$, and
$\tilde{S}^{(0)}(k)\equiv 1$. These functions are related to each
other by a recursion
\begin{equation}\label{recsn}
{\tilde S}^{(l)}(k)=\frac{1-(1+2i\lambda_1 k)\,e^{2i k}{\tilde S
}^{(l-1)}(k)}{1-2i\lambda_1 k -e^{2i k}{\tilde S}^{(l-1)}(k)}\,.
\end{equation}

Using as before the ansatz $z_m=m\pi+\zeta_m$  in the resonance
region $\lambda_{1,{\textsc{l}}}|z|\approx \lambda_{1,{\textsc{l}}}
|m|\pi\ll 1$, we arrive at an algebraic equation
$\mathcal{P}^{({\textsc{l}})}(e^{2i\zeta_m})=0$, with
$\mathcal{P}(w)$ being a polynomial of the $L$th power with respect
to the argument $w=e^{2i\zeta_m}$. This equation gives a bunch of $L$
close complex poles of the $S$ function well separated from all the
other poles.

The corresponding $R$ function is readily found from
Eq.~(\ref{explamS}) to be
\begin{equation}\label{explamR}
R(k) = -\frac{2}{1/\lambda_{\textsc{l}}
k+\cot(\delta_{\infty}+ka)}\,.
\end{equation}
In the limit $\lambda_{\textsc{l}}\rightarrow 0$ (closed interior) we
have $R(k)\equiv 0$ and $\tilde{S}(k)\equiv 1$, so that only the
immediate reflection at the point $a$ survives. Obviously, the
spectrum of the poles of the function $R(k)$ is determined by the
equation
\begin{equation}\label{tilRspec}
\sin\left(\delta_{\infty}+ka\right)+\lambda_{\textsc{l}}
k\cos\left(\delta_{\infty}+ka\right)=0
\end{equation}
which should be compared with Eq.!(\ref{poltilR1}). It is easy to
check that the spectrum exactly coincides with that of the wave
numbers of the internal problem for the potential
$U_{\infty}(r)=\kappa_1\sum_{l=1}^{L-1}\delta(r-l)$ with the BC
(\ref{cNint}). This potential is perfectly transparent on the
separation radius $a=L$ where the latter condition is fixed. This
directly follows from the expression
\begin{equation}\label{intwf}
u^0(r) = \sin(kr)
  \textstyle
-\frac{1}{2}\,T_{\infty}(k)\,e^{ikr}=\frac{i}{2}\left[e^{-ikr}-
S_{\infty}(k)\,e^{ikr}\right]
\end{equation}
for the wave function in the region $(L-1)<r\leq a=L$.

\subsection{Two $\delta$ barriers; resonance trapping}

Now we will use the formulaes just found to analyze as an
illustrative example resonances in the double-well
$\delta$-potential, $L=2$. The equation for the poles of the $S$
function (\ref{explamS}) looks in this case like
\begin{eqnarray}\label{deq2}
& (1+2i\lambda_1 z)\,e^{4iz}-2(1-2i\lambda_2 z)\,e^{2iz} \nonumber\\
& +(1-2i\lambda_1 z)(1-2i\lambda_2 z)=0\,,
\end{eqnarray}
[cf. Eq.~(\ref{deq})]. Again, the ansatz $z_m=m\pi+\zeta_m$ with
$|\Ree\zeta_m|\leq \pi/2$ is valid. Supposing also the integer number
$m$ to be appreciably large, we arrive for each fixed $m$ at a couple
of closed solutions [compare with (\ref{appzeta})]
\begin{eqnarray}\label{appzeta2}
\zeta_m &\approx& -\frac{i}{2}\ln
\left\{1-i\frac{(2\lambda_1+\lambda_2)\,m\pi}{1+2i\lambda_1m\pi} \right.\\
\nonumber & &\times
\left.\left[1\pm\sqrt{1-\frac{4\lambda_1\lambda_2}{(2\lambda_1+\lambda_2)^2}
\left(1+2i\lambda_1m\pi\right)}\right]\right\} \,.
\end{eqnarray}
It is worth noting that within the approximation adopted the sum of
the imaginary parts
$\Imm(\zeta^{+}_m+\zeta^{-}_m)=-\frac{1}{2}\ln|Z^{+}_m Z^{-}_m|=
-\frac{1}{2}\ln\sqrt{1+4\lambda^2_2(m\pi)^2}$, where $Z^{\pm}_m$
stand for the arguments of the logarithm in Eq.~(\ref{appzeta2}),
{\it does not depend} on the penetrability constant $\lambda_1$ of
the interior barrier. Indeed, simple transformations show that
$Z^{+}_m Z^{-}_m=\left[(1-2i\lambda_1 m\pi)/(1+2i\lambda_1
m\pi)\right](1-2i\lambda_2 m\pi)$.

We fix now the number $m\gg 1$ within the resonance domain,
$\lambda_{1,2}|m\pi|\ll 1$, and consider the doublet of poles close
to this point. Keeping the terms of the two first orders of magnitude
we find
\begin{eqnarray}\label{doublet}
\zeta_m^{\pm}&\approx& -\frac{1}{2} \left(2\lambda_1+
\lambda_2\pm\sqrt{4\lambda_1^2+\lambda^2_2}\right)m\pi
\nonumber \\
&&-\frac{i}{2}\,\lambda^2_2
\left(1\pm\frac{\lambda_2}{\sqrt{4\lambda_1^2+
\lambda^2_2}}\right)(m\pi)^2\,.
\end{eqnarray}
In the resonance region, the splitting $|\Delta z_m|=|\Delta
\zeta_m|\approx
\sqrt{4\lambda_1^2+\lambda^2_2}$ within a given doublet is much
smaller than
the distance between adjacent doublets which is $\sim \pi$.

Equation (\ref{explamR}) reads in the case $L=2$
\begin{equation}\label{tilR2}
R(k) = -\frac{2\lambda_2 k\,(\sin^2 k+\lambda_1 k\sin 2k) }{ \sin^2
k+\frac{1}{2}(2\lambda_1+\lambda_2)k\sin 2k+\lambda_1\lambda_2 k^2\cos
2k}\,.
\end{equation}
For the $m$th resonance doublet $k_m=m\pi+\delta k_m$, and the small
shift $\delta k_m$ satisfies the quadratic equation
\begin{equation}\label{polR2}
(\delta k_m)^2 +(2\lambda_1+\lambda_2)(m\pi)\,\delta
k_m+\lambda_1\lambda_2
(m\pi)^2=0 \,,
\end{equation}
giving immediately $\delta k_m^{\pm}=\Ree\zeta_m^{\pm}$, with
$\zeta_m^{\pm}$ from (\ref{doublet}).

This convinces us that each resonance doublet with a good accuracy
can be considered independently of the other poles. Near a given
$k_m=m\pi$ within the resonance region the function $R(m\pi+\delta
k)\equiv R_m(k)$ is decomposed into a sum of two partial fractions
\begin{equation}\label{pexp2}
R_m(k)=\frac{\gamma^{+}_m}{k-k^{+}_m}+ \frac{\gamma^{-}_m}{k-k^{-}_m}
\,,
\end{equation}
the residues being found to be equal to
$\gamma_m^{\pm}=-2\Imm{\zeta_m^{\pm}}$.

To pass from the $k$ to the energy plane, the term $R_{-m}(k)$ must
be added, which yields
\begin{equation}\label{R2ofE}
R_m(E) = \frac{\Gamma_m^{+}}{E-\varepsilon_m^{+}}+
\frac{\Gamma_m^{-}}{E-\varepsilon_m^{-}}
 = A_m^T\frac{1}{E-\varepsilon^{\textsc{(n)}}_m}A_m\,.
\end{equation}
The two-dimensional vector
$A^T_m$=$(\sqrt{\Gamma_m^{+}},\sqrt{\Gamma_m^{-}})$ and matrix
$\varepsilon^{\textsc{(n)}}_m$=$\textrm{diag\,}(\varepsilon^{+}_m,
\varepsilon^{-}_m)$ of the internal levels
$\varepsilon^{\pm}_m$=$(k^{\pm}_m)^2$ of the doublet have been
introduced on the last step.

The corresponding representation of the scattering amplitude near the
scattering energy $E\approx k_m^2$ is as follows:
\begin{equation}\label{S2ofE}
{\tilde T}_m(E)=A_m^T\frac{1}{E-\left(\HeffN\right)_m}\,A_m\,.
\end{equation}
In these formulas $\Gamma_m^{\pm}=2\sqrt{E}\,\gamma_m^{\pm}\approx
2\pi m\gamma_m^{\pm}$, i.e., we have neglected the change of the
scattering energy $E$ within the doublet considered. The difference
between $k_m^{\pm}$ contributes only at higher orders. This is
contrary to the case of widths from different doublets when $k_m$
differ from each other already in the zero order in $\lambda_{1,2}$.
In such an approximation, both functions $R_m(E)$ and
$\tilde{T}_m(E)$ become meromorphic in the complex energy plane.

By the very construction, the complex eigenvalues of the
two-dimensional {\it energy-independent} symmetric matrix
\begin{equation}\label{effH2}
\HeffN = \varepsilon^{\textsc{(n)}}-\frac{i}{2}\,{\tilde A} {\tilde
A}^T
\end{equation}
coincide with the complex energies of the two resonances of the
doublet considered. Henceforth, we drop the index $m$ of the doublet
to avoid too bulky notation. The energies $\varepsilon^{\pm}$ are the
levels of the internal problem for the potential
$U^{(2)}_{\infty}(r)=\kappa_1\delta(r-a)$ with the BC (\ref{cNint})
whereas the amplitudes are given by Eq.~(\ref{lamA}), with the wave
number $k$ substituted by $m\pi$. Just the {\it energy-independent}
matrix (\ref{effH2}) is naturally interpreted as an effective
non-Hermitian Hamiltonian. The notion of the effective Hamiltonian is
valid, however, only within a fixed doublet. Similarly, in the case
$L=1$ the ``effective Hamiltonian'' coincides with the corresponding
diagonal matrix element of the formal operator $\mathcal{H}$
constructed in Sec.~III.

\begin{figure}
\unitlength 1cm
\begin{picture}(8,9.5)
\epsfxsize 7.5cm \put(0.35,-0.2){\epsfbox{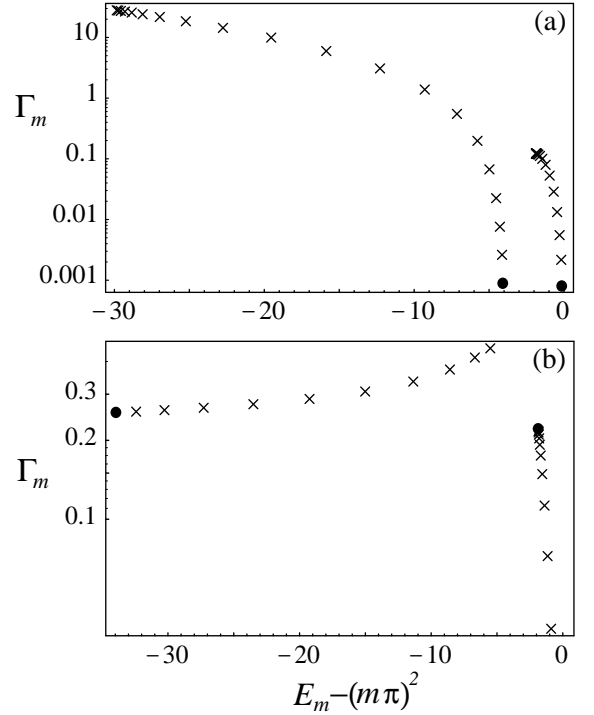} }
\end{picture}
\caption{Widths (in log scale) versus energies for a doublet of
resonances with $m$=10 as (a) $\lambda_1/\lambda_2$ is varied from
0.1 to 10 (marked with $\bullet$) at fixed $\lambda_1$=0.001; (b)
$\lambda_1/\lambda_2$ is varied from 1 to 10 (marked with $\bullet$)
at fixed $\lambda_2$=0.001. }
\end{figure}
As we have already mentioned above, the total width $\Gamma
=\Gamma_{+}+\Gamma_{-}\approx4\lambda^2_2 (m\pi)^3$ of the doublet
remains constant as long as the openness of the system is fixed. At
the same time, the individual widths depend also on the ratio
$\lambda_1/\lambda_2$. A {\it remarkable redistribution} of the total
width between members of the doublet takes place when this parameter
changes. When the system is almost closed, $\lambda_2\ll\lambda_1$,
\begin{equation}\label{2<1}
\begin{array}{l}
E_{+}\approx (1-4\lambda_1-\lambda_2)\,(m\pi)^2\,,\\[1ex] %
E_{-}\approx (1-\lambda_2)(m\pi)^2 \,,
\end{array}
\quad \Gamma_{\pm}\approx 2\lambda^2_2 (m\pi)^3\,.
\end{equation}
Both levels have similar widths in this case. However, in the
opposite limit $\lambda_2\gg\lambda_1$ the state that exists in the
outer well appropriates almost the whole total width
\begin{eqnarray}\label{2>1}
E_{+}\approx (1-2\lambda_2-2\lambda_1)(m\pi)^2\,,  &&
\Gamma_{+}\approx 4(\lambda^2_2-\lambda_1^2)(m\pi)^3\,, \nonumber\\
E_{-}\approx
(1-2\lambda_1)(m\pi)^2 \,, &&
\Gamma_{-}\approx 4\lambda_1^2(m\pi)^3\,.
\end{eqnarray}
Simultaneously, the energy $E_+$ of the broader resonance gets
strongly displaced due to the coupling to the energy continuum. The
other resonance turns out to be trapped in the inner well. Figure 3
illustrates the behavior just described. For an {\it arbitrary} ratio
$\lambda_1/\lambda_2$, the complex energies of the two resonances of
a doublet are separated by a distance which is large on the scale of
the total width \cite{degeneracy}. For this reason, the off-diagonal
matrix elements of the anti-Hermitian part of the effective
Hamiltonian (\ref{effH2}) give only corrections of higher order and
can therefore be neglected. In spite of the nontrivial behavior of
the complex levels just described the resonances may still be
considered separately of each other within the given accuracy.

Extension to the general case of an arbitrary number $L$ of the
barriers is now straightforward. The effective Hamiltonian appears as
a $L\times L$ block of the operator $\mathcal{H}$, which describes a
fixed bunch of $L$ close resonance states. It is important that
energy dependence can be fully neglected within such a bunch.

\subsection{Internal problem with Dirichlet BC}

The internal problem with the BC (\ref{cNint}) does not correspond to
a finite motion of a quantum particle even when the coupling matrix
elements are neglected. Indeed, the internal solution is sensitive to
the penetrability of the outer barrier. Meanwhile, in numerous
applications the intuitively most attractive convention, which goes
back to the textbook \cite{Mahaux1969}, is adopted, using as an
internal basis the states of a closed counterpart of the system under
consideration. Such a counterpart can hardly be defined uniquely but,
in our case, it is natural to fix it by choosing Dirichlet BC at the
point $r=a$. This corresponds to the internal problem with an
infinitely hard wall put at the point $r=a$. The formalism of Sec.~II
gives then
\begin{subequations}\label{DlamS}
\begin{eqnarray}
S(k) &=& \frac{1-\frac{i}{\lambda_{\textsc{l}}k}
[1+\lambda_{\textsc{l}}{\GintD}''(a,a)] }{
1+\frac{i}{\lambda_{\textsc{l}}k}
[1+\lambda_{\textsc{l}}{\GintD}''(a,a)]}
S_0^{\textsc{(n)}} \\
\label{DlamS0}
&=& \frac{1+\frac{\lambda_{\textsc{l}}}{1+i\lambda_{\textsc{l}} k
}{\GintD}''(a,a)}{1+\frac{\lambda_{\textsc{l}}}
{1-i\lambda_{\textsc{l}} k
}{\GintD}''(a,a)}\,\frac{1+i\lambda_{\textsc{l}}
k}{1-i\lambda_{\textsc{l}}
k}\,S_0^{\textsc{(d)}}.\mbox{\quad}
\end{eqnarray}
\end{subequations}
Comparison with the closed solution (\ref{lamS}) and (\ref{explamS})
shows that
\begin{equation}\label{DlamR}
\cot\left(\delta_{\infty}+ka\right) = \frac{1}{k}{\GintD}''(a,a) =
\frac{1}{k}\, \sum_n
\frac{{v^0_n}'(a_{-})\,{v^0_n}'(a)}{k^2-\varepsilon_n^{\textsc{(d)}}},
\end{equation}
which implies that the spectrum of the internal problem is given by
the equation $\sin\left(\delta_{\infty}+ka\right)=0$ instead of
Eq.~(\ref{tilRspec}). Unlike the eigenvalues
$\varepsilon^{\textsc{(n)}}_n$, the levels
$\varepsilon^{\textsc{(d)}}_n$ {\it do not depend} on
$\lambda_{\textsc{l}}$.

Similar to Eqs.~(\ref{lamS})--(\ref{lamT}), we have
$S(k)=\bar{S}(k)S_0(k)$, with $S_0$ from Eq.~(\ref{vlev}), and
$\bar{S}(k)=1-i{\bar A}^T[k^2-\mathcal{H}^{\textsc{(d)}}]^{-1}{\bar
A}$, where
\begin{eqnarray}\label{lamcalHD}
\mathcal{H}^{\textsc{(d)}}_{mn} &=&
\varepsilon_m^{\textsc{(d)}}\delta_{mn} -
\frac{\lambda_{\textsc{l}}}{1-i\lambda_{\textsc{l}}k}\,
{v^0_m}'(a)\,{v ^0_n}'(a) \nonumber \\
&=&
\varepsilon_m^{\textsc{(d)}}\delta_{mn}-\frac{1}{2\lambda_{\textsc{l}}
k}\, {\bar A}_m {\bar A}_n-\frac{i}{2}\,{\bar A}_m {\bar A}_n
\end{eqnarray}
and
\begin{equation}\label{barA}
{\bar A}_n = \lambda_{\textsc{l}}\sqrt{\frac{2k }{
1+\lambda^2_{\textsc{l}} k^2}}\,{v^0_n}'(a)\,.
\end{equation}
This should be compared with Eq.~(\ref{lamA}) and (\ref{lamcalH}).
Not only widths but also Hermitian shifts result from the coupling to
the external region this time. Indeed, in the resonance region
$\lambda_{\textsc{l}}k\ll 1$ the phases that come from the first two
factors in Eq.~(\ref{DlamS0}) are equal approximately to
\begin{equation}\nonumber
{\bar\delta}(k)\approx -\lambda_{\textsc{l}}^2
k\frac{{\GintD}''(a,a)}{1+\lambda_{\textsc{l}}{\GintD}''(a,a)}\,;
\quad
\delta_{\lambda_{\textsc{l}}}(k)\approx \lambda_{\textsc{l}} k\,.
\end{equation}
The two contributions perfectly compensate each other at the points
$k_n^2=\varepsilon_n^{\textsc{(d)}}$. Within a bunch of resonance
levels the eigenvalues of the Hermitian part of $\HeffD$ coincide in
the main approximation with the corresponding levels
$\varepsilon_n^{\textsc{(n)}}$. In general, the connection
${\tilde\delta} (k)={\bar\delta}(k)+\delta_{\lambda_{\textsc{l}}}(k)$
holds for arbitrary $k$.

Let us first return to the case $L=2$, for a moment. The amplitudes
(\ref{barA}) for the $m$th doublet are  easily calculated to be in
the main approximation ${\bar A}_{\pm}\approx \pm
\sqrt{2}\lambda_2\,(m\pi)^{3/2}$, when the corresponding internal
levels are
$\varepsilon^{\textsc{(d)}}_{+}\approx(1-4\lambda_1)\,(m\pi)^2$ and
$\varepsilon^{\textsc{(d)}}_{-}=(m\pi)^2$. Therefore, the diagonal
elements of the effective Hamiltonian of the doublet are equal to
\begin{equation}\label{calE0D2}
\mathcal{E}^{(0)}_{\pm} \approx \left\{\begin{array}{l}
\left(1-4\lambda_1-\lambda_2\right)\,(m\pi)^2-i\lambda_2^2(m\pi)^3,
\\[1ex]
\left(1-\lambda_2\right)\,(m\pi)^2-i\lambda_2^2(m\pi)^3\,,
\end{array}\right.
\end{equation}
which exactly coincide with Eq.~(\ref{2<1}). Off-diagonal elements of
$\HeffD$ can be neglected if $\lambda_2\ll\lambda_1$. These elements
become important, however, when the coupling to the continuum becomes
strong, $\lambda_2\gtrsim\lambda_1$. The two resonances
(\ref{calE0D2}) interfere in this case. Diagonalization of the
$2\times 2$-matrix $\HeffD$ gives now
\begin{eqnarray}\label{calED2}
\mathcal{E}_{\pm} &\approx & \left[1-2\lambda_1-\lambda_2
\mp\sqrt{4\lambda_1^2+\lambda_2^2}
\right]\,(m\pi)^2 \nonumber\\
&& -i\lambda_2^2\left(1\pm\frac{\lambda_2}
{\sqrt{4\lambda_1^2+\lambda_2^2}}\right)(m\pi)^3.
\end{eqnarray}
In the limit $\lambda_2\gg\lambda_1$ this reduces to (\ref{2>1}).

For an arbitrary number $L$ of barriers, the $m$th bunch of $L$ close
resonances is described by the $L\times L$-submatrix $\HeffD$ of the
infinite matrix (\ref{lamcalHD}), with the wave number $k$ being
substituted by $k_m\approx m\pi$. The absolute values of the coupling
amplitudes within such a resonance bunch are estimated as $|{\bar
A}_n|\approx 2\lambda_{\textsc{l}} (m\pi)^{3/2}/\sqrt{L}$. The total
{\it collective} width $\Gamma^{(m)}_c=\sum_{l=1}^L {\bar
A}_l^2\equiv {\bf\bar A}^2 \approx 4\lambda^2_{\textsc{l}} (m\pi)^3$
of the bunch, which is determined by the trace of the anti-Hermitian
part of $\HeffD$, characterizes the openness of the system and does
not depend on the penetrability constant $\lambda_1$ of the internal
barriers. The same is valid in the main approximation regarding the
collective real energy displacement $\delta\epsilon^{(m)}_c\equiv
2\lambda_{\textsc{l}} (m\pi)^2 \approx
\Gamma_c^{(m)}/2\lambda_{\textsc{l}}m\pi$. When this displacement is
small in comparison with the total energy spread of the internal
levels,
$\delta\epsilon^{(m)}_c\ll\Delta^{(m)}(\varepsilon^{\textsc{(d)}})$,
the bunch consists of $L$ independent narrow resonances
\begin{equation}\label{sDlamS}
\mathcal{E}^{(m)}_n \approx \varepsilon^{\textsc{(d)}}_n -
\frac{1}{L}\,\delta\epsilon^{(m)}_c-
\frac{i}{2}\frac{1}{L}\,\Gamma_c^{(m)}\,.
\end{equation}
Note that the
Hermitian shift does not influence the level spacings but changes
only the position of the bunch.

Under the opposite condition, $\delta\epsilon^{(m)}_c\gg \Delta^{(m)}
(\varepsilon^{\textsc{(d)}})$, the doorway basis in which the
interaction matrix ${\bar A}{\bar A}^T$ is diagonal becomes more
adequate \cite{Sokolov1992} than the basis of the internal problem
used up to now (see also \cite{Sokolov1997i}). The latter matrix is
diagonalized by an orthogonal transformation
$\eta=\left(\mbox{\boldmath$\eta$}^{(1)}\;\mbox{\boldmath$\eta$}^{(2)}
\;... \mbox{\boldmath$\eta$}^{\textsc{(l)}}\right)$, where each entry
$\mbox{\boldmath$\eta$}^{(l)}$ is a real vector in the
$L$-dimensional part of the total Hilbert space, which corresponds to
the considered bunch. Due to the factorized structure, the
interaction matrix possesses the only nonzero eigenvalue
$\Gamma_c^{(m)}={\bf\bar A}^2$, which belongs to the eigenvector
$\mbox{\boldmath$\eta$}^{(1)}={\bf a}\equiv {\bf\bar
A}/\sqrt{{\bf\bar A}^2}$. As a result, the effective Hamiltonian
obtains in the doorway basis the form (we drop below the index $m$ of
the bunch)
\begin{equation}\label{DRH}
\HeffD =
\left(\begin{array}{cc}
\epsilon_c-\frac{i}{2}\,\Gamma_c & {\bf h}^{\textsc{t}} \\
{\bf h}  & {\hat\epsilon} \end{array}\right)\,.
\end{equation}
The following notation has been used here ($\mu$=2,3,...,$L$):
\begin{equation}\label{collev}
\begin{array}{l}
\epsilon_c=\sum_l\varepsilon_l^{\textsc{(d)}}a_l^2-2\lambda_{\textsc{l}}
(m\pi)^2=\langle\varepsilon_l^{\textsc{(d)}}\rangle-\delta\epsilon_c\,,\\[1ex]
h_{\mu}=\sum_l\varepsilon_l^{\textsc{(d)}}\,a_l\,\eta_l^{(\mu)}\,,
\end{array}
\end{equation}
with the $L-1$ vectors $\mbox{\boldmath$\eta$}^{(\mu)}$ and elements
$\epsilon_{\mu}$ of the diagonal matrix ${\hat\epsilon}$ being
defined by the eigenvalue problem
\begin{equation}\label{subeta}
\sum_n\varepsilon_n^{\textsc{(d)}}\,\eta_n^{(\mu)}\,\eta_n^{(\nu)}=
\epsilon_{\mu}\,\delta_{\mu\nu}
\end{equation}
in the $(L-1)$-dimensional subspace orthogonal to the vector ${\bf
a}$. The quantity $\langle\varepsilon_l^{\textsc{(d)}}\rangle$ is the
weighted mean position of the internal levels. The non-Hermitian
Hamiltonian (\ref{DRH}) describes a wide {\it doorway} resonance with
width $\Gamma_c$ displaced from the bunch by the distance
$\delta\epsilon_c$ and coupled to a background of $L-1$ stable states
by means of the matrix elements $h_{\mu}$. Only due to this
interaction do such states get access to the continuum via the
doorway state existing in the outer well.

With the help of the completeness condition
\begin{equation}\label{comp}
\sum_{\mu}\,\eta^{(\mu)}_l\,\eta^{(\mu)}_{l'}=\delta_{ll'}-a_l\,a_{l'}
\end{equation}
one finds from Eq.~(\ref{subeta})
\begin{equation}
\eta^{(\mu)}_l = -h_{\mu}
\frac{1}{\epsilon_{\mu}-\varepsilon^{\textsc{(d)}}_l}\,a_l\,.
\end{equation}
The orthogonality condition ${\bf
a}\cdot\mbox{\boldmath$\eta$}^{(\mu)}=0$ immediately gives the
equation
\begin{equation}
\sum_l\frac{a_l^2}{\epsilon_{\mu}-\varepsilon^{\textsc{(d)}}_l}=0
\end{equation}
for the new positions of the stable levels. This equation shows that
each new level $\epsilon_{\mu}$ lies between two neighboring old ones
$\varepsilon^{\textsc{(d)}}_l$ and therefore is shifted with respect
to the latter only by a distance comparable with the initial mean
level spacing. This is much smaller than the displacement
$\delta\epsilon_c$ of the collective level $\epsilon_c$.

The interaction ${\bf h}$ mixes these states and forms $L$ final
resonance states. The complex energies of the exact states, i.e. the
eigenvalues of the effective Hamiltonian (\ref{DRH}), satisfy the
secular equation
\begin{equation}\label{calEND}
{\cal
E}=\epsilon_c-\frac{i}{2}\,\Gamma_c+\sum_{\mu}\frac{h_{\mu}^2}{{\cal
E}-\epsilon_{\mu}}\,.
\end{equation}
In particular, for the collective doorway state one obtains from this
equation
\begin{equation}
\mathcal{E}_{coll} \approx
\langle\varepsilon_l^{\textsc{(d)}}\rangle-
\left[1-\frac{{\bf h}^2}{(\delta\epsilon_c)^2}\right]
\delta\epsilon_c-\frac{i}{2}\,
\left[1-
\frac{{\bf h}^2}{(\delta\epsilon_c)^2}\right]\Gamma_c\,.
\end{equation}
Equations (\ref{collev}) and (\ref{comp}) allow us to express the
square length ${\bf h}^2$ of the mixing vector  in terms of the
variance of the non-perturbed internal levels,
\begin{eqnarray}
{\bf h}^2 &=& \sum_l (\varepsilon^{\textsc{(d)}}_l)^2\,
a_l^2-\left(\sum_l\varepsilon^{\textsc{(d)}}_l\,
a_l^2\right) \nonumber\\
&=& \langle\left(\varepsilon^{\textsc{(d)}}-
\langle\varepsilon^{\textsc{(d)}}\rangle\right)^2\rangle
= [\Delta
\varepsilon^{\textsc{(d)}}]^2\,.
\end{eqnarray}
Thus, interaction is weak in the doorway basis under the condition
$\delta\epsilon_c\gg \Delta\varepsilon^{\textsc{(d)}}$ and the doorway state
keeps almost the whole energy displacement and width. All other states are
trapped in the interior region and share small portions
$\sim[\Delta\varepsilon^{\textsc{(d)}}/\delta\epsilon_c]^2$ of the collective
displacement and width in accordance with the group velocity
\cite{Sokolov1992} . The solution just described has formally very much in
common with the  schematic model of the so-called nuclear dipole giant
resonance developed in \cite{Sokolov1990,Sokolov1997i}.

\section{Conclusion}

In this paper we analyzed the relevance of the concept of the
non-Hermitian effective Hamiltonian in finite-range potential
scattering. Single-channel $s$-wave resonance scattering is
considered as an example. The scattering function $S(k)$ is
meromorphic in this case, i.e., has only isolated poles in the
complex plane of the wave numbers. The number of poles is, generally,
infinite but only a finite part of them can be interpreted as
resonances.

We first presented (Secs. II and III) a consistent self-adjoint
formulation of the scattering problem, which is based on separation
of the configuration space into internal and external segments. In
this way, the $S(k)$ function is represented in terms of a
non-Hermitian energy-dependent operator $\mathcal{H}(k)$. There
exists a wide freedom in choosing the radius $a$ of separation as
well as the boundary conditions at this point. Different choices
yield different explicit forms of this operator together with the
$S(k)$ and $R(k)$ functions. This can, in particular, come out
strongly if one truncates the matrix $\mathcal{H}$ to calculate
numerically the positions of the poles of $S$ function in the complex
$k$ plane. Nevertheless, the true complex poles of the $S$ function
and this function itself depend, as we explicitly demonstrate,
neither on the BC nor on the radius $a$.

Although all choices are formally allowed, this does not mean that
they are equally adequate from the physical point of view. For
instance, a fictitious immediate reflection takes place at the
(arbitrarily chosen) point of separation $a$. The artificial
separation of the phase of this reflection results, in turn, in the
appearance of an infinite number of remote poles of the function
$R(k)$, which describe the smooth contribution from the interior
region $r<a$. The phase of the immediate reflection must be
compensated by the contributions of such poles for the radius $a$ to
disappear finally from the scattering amplitude. This can cause
unjustified complications in intermediate stages of calculations. We
argue that the harm is minimized if the separation radius matches an
outer potential barrier when it is strong enough to make immediate
reflection at this point quite probable.

The notion of the non-Hermitian effective Hamiltonian first
introduced in the theory of resonance nuclear reactions emerges when
a group of very close resonance states strongly overlap and
interfere. We stress that, in contrast to the nuclear reactions, in
the cases of the potential resonance scattering usually discussed in
the literature the density of the energy spectrum is too low and the
ordinary Breit-Wigner approximation of isolated resonances usually
suffices. The complex energy of the $m$th resonance which dominates
in the scattering amplitudes near the scattering energy $E\approx
|z_m|^2$ is well approximated by the diagonal matrix element
$\mathcal{H}_{mm}(|z_m|)$. There is no room for an effective
Hamiltonian in this case or, more strictly, it is embodied by the
$1\times 1$ ``matrix'' $\mathcal{H}_{mm}$. We emphasize that the
energy dependent operator $\mathcal{H}(k)$ should not be confused
with the effective Hamiltonian. Energy dependent eigenvalues of this
operator are not, generally, in one-to-one correspondence with the
complex energies of the actual resonance states.

Overlap and interference of the resonances can become possible in the
cases when the energy spectrum has a band structure. As an example of
this kind we investigate a periodically disposed chain of a finite
number $L$ of radial $\delta$ barriers. All $S$, $K$ and $R$
functions are found in closed forms in terms of an $L\times L$
$k$-dependent matrix propagator. This allows us to study in detail
all analytical properties in the complex $k$ plane and to verify the
correspondence with the projection formalism used. There exist a
finite number of separated bands of close resonances. Within the
$m$th band which lie near the scattering energy $E\approx |z_m|^2$
one can neglect all smooth variations with $k$. In this
approximation, the mentioned propagator proved to coincide with the
resolvent of the $L\times L$ block ${\cal H}_{ll'}(|z_m|)$ of the
matrix ${\cal H}^{\textsc{(n)}}_{mn}(k)$ {\it taken at the fixed
value} $k=|z_m|$. Just this matrix plays the role of the effective
Hamiltonian of the system in the energy interval within the band.

Different choices of BCs yield different patterns of the resonance
interference. In particular, the spectrum of the poles of the $R$
function is exactly reproduced in the framework of the projection
technique with the BC (\ref{cNint}) of Neumann type fixed at the
position $a=L$ of the outer barrier. The corresponding energy levels
depend on the penetrability constant $\lambda_{\textsc{l}}$ of the
outer barrier. Shifts of the levels due to the coupling to the
continuum are included in this case from the very beginning.

Utilization of the Dirichlet BC for the intrinsic motion gives
another but yet equivalent formulation of the scattering problem
considered. The internal problem in this case fixes a closed
counterpart of the open system under consideration. This enables us
to investigate the change of the regime of the internal motion of the
system as its openness grows. The interaction via the continuum
shifts the original $\lambda_{\textsc{l}}$-independent levels of the
internal motion along both real and imaginary axes. A transition is
explicitly demonstrated from the bunch of $L$ similar narrow
resonances to the formation of a relatively broad resonance strongly
shifted with respect to the band, which exists in the outer well. The
other resonances turn out to be trapped in the inner part of the
system. This is  quite similar to the results of  earlier
investigations which rely upon the notion of the non-Hermitian
effective Hamiltonian of an open system.

\section*{Acknowledgments}

We are grateful to Y.V. Fyodorov for discussions on Feshbach's
projector technique in an early stage of this work, to H. Schanz and
V. Zelevinsky for interesting discussions, and to G. Hackenbroich and
C. Viviescas for their interest in the work and useful conversations.
One of us (V.V.S.) greatly appreciates the generous hospitality of
the Max-Planck Institute for the Complex Systems extended to him
during his stay in Dresden. The financial support by RFBR Grant No.
03-02-16151 (D.V.S. and V.V.S.) and SFB 237 ``Unordnung and grosse
Fluktuationen'' (D.V.S. and H.J.S.) is acknowledged with thanks.


\begin{thebibliography}{29}
\bibitem{convergence}
 As usual, factors $e^{-\alpha n}$ with infinitesimal $\alpha$
 should be introduced to guarantee the absolute convergence.
\bibitem{barrier}
Let us recall that the penetrability of a $\delta-$barrier
$D(k)=(\lambda k)^2/[1+(\lambda k)^2]$  is small,
$D(k)\approx(\lambda k)^2$, if $\lambda k\ll 1$, and goes to one
under the opposite condition.
\end{thebibliography}

\end{document}